\pdfoutput=1

\documentclass[twocolumn,english,prl]{revtex4-1}
\usepackage[latin9]{inputenc}
\usepackage{amsmath}
\usepackage{graphicx}

\makeatletter
\usepackage{upgreek}
\usepackage{mathtools}
\usepackage{verbatim}

\renewcommand{\vec}{\boldsymbol}

\newcommand\asteq{\stackrel{\ast}{=}}

\usepackage{babel}

\usepackage{braket}

\makeatother

\usepackage{babel}
\begin{document}

\title{Light emission by accelerated electric, toroidal and anapole dipolar
sources}

\author{V. Savinov}

\affiliation{Optoelectronics Research Centre and Centre for Photonic Metamaterials,
University of Southampton, Southampton SO17 1BJ, UK }
\begin{abstract}
Emission of electromagnetic radiation by accelerated particles with electric, toroidal and anapole dipole moments is analyzed. It is shown that ellipticity of the emitted light can be used to differentiate between electric and toroidal dipole sources, and that anapoles, elementary neutral non-radiating configurations, which consist of electric and toroidal dipoles, can emit light under uniform acceleration. The existence of non-radiating configurations in electrodynamics implies that it is impossible to fully determine the internal makeup of the emitter given only the distribution of the emitted light. Here we demonstrate that there is a loop-hole in this `inverse source problem'. Our results imply that there may be a whole range of new phenomena to be discovered by studying the electromagnetic response of matter under acceleration.
\end{abstract}
\maketitle

\section{Introduction\label{sec:Introduction}}

Electromagnetic radiation is produced by oscillating and accelerating
charges and currents. The converse, however, is not true. There exists
a wide class of configurations composed of oscillating charges and
currents which emit no electromagnetic fields. These are known as
the nonradiating configurations \citep{Devaney73}. The nonradiating
configurations are not merely a mathematical curiosity. Their existence
directly implies that it is impossible to deduce the internal composition
of a charge-current configuration given only information about the
fields it emits. In practice, this means that if, given a field distribution,
a suitable source charge-current configuration can be found, it will
not be unique, since one can always add a nonradiating configuration
to it without changing the emitted field. This feature of Maxwell's
equations, also known as the inverse source problem \citep{Bleistein,Stone87,DevaneyBook12},
impacts many branches of science where light is used to interrogate
distant or otherwise inaccessible objects, from medical imaging, to
astronomy and radar science \citep{DevaneyBook12}. Here we will show
that neutral nonradiating configurations can, in fact, be made to
radiate by using acceleration. This is a loop-hole in the inverse
source problem. Unlike other techniques for characterizing localized
oscillating charge-current configurations, such as near-field microscopy
for example \citep{NovotnyBook12}, our solution is universal since
it only relies on properties of Minkowski spacetime and Maxwell's
Equations in vacuum. 


\begin{figure}
\includegraphics{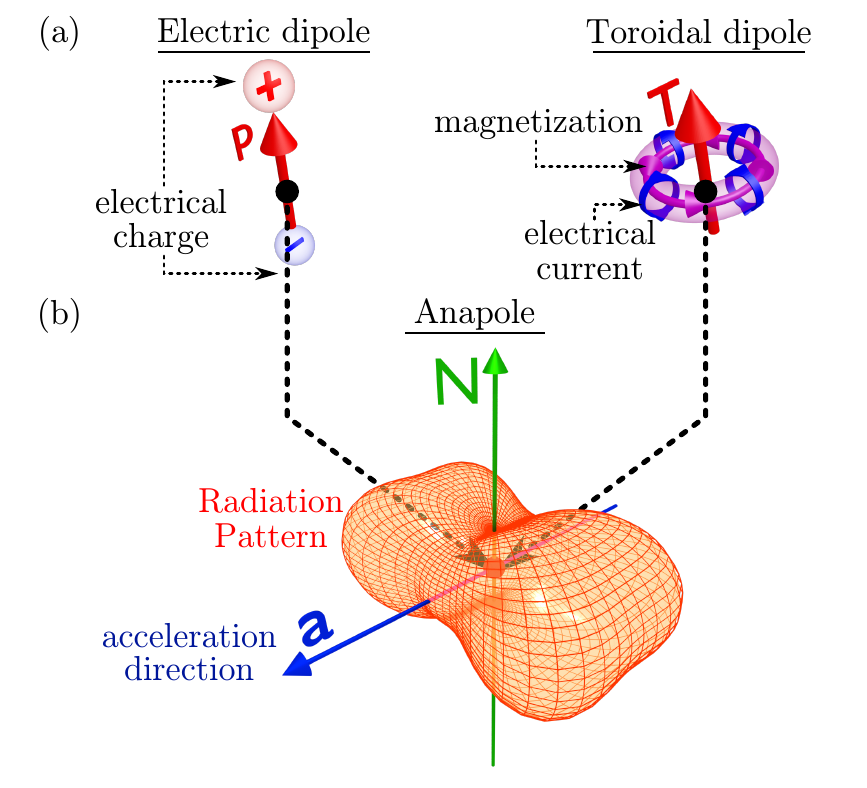}

\caption{\textbf{Radiation pattern of a point-like nonradiating configuration
(anapole) subjected to acceleration.} \textbf{(a)}~Electric and toroidal
dipoles that together make up the anapole. The electric dipole ($\vec{p}$)
corresponds to two separated opposite charges. The toroidal dipole
($\vec{T}$) corresponds to a loop of magnetization, or, equivalently,
to current flowing along the meridians of a torus. \textbf{(b)}~Radiation
pattern (radius represents power per solid angle) of a dynamic anapole
with moment $\vec{N}$, accelerating with constant acceleration $\vec{a}$.}

\label{fig:AnapoleGeneral}
\end{figure}

The key to understanding the general properties of nonradiating configurations
is to consider their elementary building blocks - anapoles. Dynamic
anapoles are point-like nonradiating configurations that consist of
co-positioned and co-aligned electric and toroidal dipoles \citep{Zeldovich57,Afa95,ToroRev16},
as shown in Fig.~\ref{fig:AnapoleGeneral}a. The electric dipole
is the usual infinitesimal electromagnetic excitation created by two
separated charges \citep{jackson}. The toroidal dipole is another
fundamental point-like excitation that can be represented as current
flowing on the surface of a torus (see Fig.~\ref{fig:AnapoleGeneral}a).
Introduced by Ya.~B.~Zeldovich in 1957 \citep{Zeldovich57} in the
context of nuclear and particle physics, toroidal dipoles have since
been discovered in a number of systems ranging from atomic nuclei
to solid state physics and artificial composite media (for a recent
review see \citep{ToroRev16}). 


An important feature of monochromatic electromagnetic radiation in
vacuum is that double-curls of both the electric and magnetic fields
are proportional to the fields themselves (i.e. $\nabla\times\nabla\times\vec{E}=k^{2}\vec{E}$
for electric field, where $k$ is the wavenumber). This symmetry leads
to the emission of electric and toroidal dipoles being identical anywhere
outside the source region \citep{ToroRev16}. Indeed, this is why
these two dipoles can be combined to create a point-like nonradiating
configuration (anapole). For a long time, dynamic, i.e. time-varying,
anapoles remained a theoretical concept. Their existence was first
demonstrated in a microwave metamaterial \citep{NonRadSciRep13},
and has since been repeated in a variety of man-made systems in various
domains of the electromagnetic spectrum \citep{Miro15,ToroRev16,NemkovExp17,Shibanuma17,Basharin17,Grinblat17}.
From the general properties of nonradiating configurations it is relatively
simple to show that any nonradiating configuration can be regarded
as a collection of anapoles \citep{Devaney73,Devaney74,Nemkov17}.
Therefore, anapoles can be used as a convenient model to understand
the general properties of all nonradiating configurations. The focus
of this work is on anapole as well as electric and toroidal dipoles
in non-inertial motion.


It is well-known that motion, and especially non-intertial motion,
changes the fields produced by charged particles \citep{jackson}.
The same applies to the fields due to neutral particles with various
multipole moments. The motion of particles with conventional electric
and magnetic multipole moments was first investigated more than 50
years ago \citep{Ellis63,Ward65,Ellis66,Monaghan68}. However the
properties of particles with toroidal dipole and anapole moments in
non-inertial motion have so far been overlooked (see App.~\ref{app_subsec:MovingToro}).
Here, for the first time, we present a rigorous treatment of radiation
produced by a uniformly accelerated neutral particle with an anapole
moment (anapole particle). In the process, we also develop the first
treatment of the radiation produced by an accelerated particle with
a toroidal dipole moment. We show that despite being nonradiating
when at rest (or in inertial motion, see App.~\ref{app_subsec:IntertialAna}),
anapole particles do produce radiation when subjected to acceleration.
By extension, this implies that all accelerated `nonradiating configurations'
do, in fact, radiate. Finally, we show that acceleration provides
a way of unambiguously differentiating between electric and toroidal
dipole excitations based solely on far-field radiation, thus overcoming
the limitation set by the inverse-source problem.


The structure of the paper is as follows. The conventional notion
of point-particles with electric, toroidal and anapole dipole moments
is generalized to a covariant description in terms of four-currents,
in Sec.~\ref{sec:FourCurrents}. The general approach of obtaining
far-field radiation from accelerating point-particles is discussed
in Sec.~\ref{sec:Radiation}. The results of these two sections are
combined to obtain the far-field radiation patterns of accelerated
dipoles, in Sec.~\ref{sec:RadiationCalc}. The results are analyzed
and summarized in Sec.~\ref{sec:Discussion}. Appendix contains additional
material on the derivations as well as a section of the standard notation
used in the paper (see App.~\ref{app_subsec:Basics}).

\section{Covariant Description of Electric, Toroidal and Anapole Dipole Point-Particles\label{sec:FourCurrents}}


This section provides the definitions of the electric, toroidal and
anapole dipole electromagnetic excitations in terms of charge and
current density, and then generalizes these definitions to four-currents
of point-particles in arbitrary motion.

As is shown in Fig.~\ref{fig:AnapoleGeneral}, an anapole corresponds
to a superposition of electric and toroidal dipoles. The charge ($\rho$)
and current ($\vec{J}$) densities of an anapole particle stationary
in the lab-frame are \citep{ToroRev16}:
\begin{gather}
\rho\left(t,\vec{r}\right)=-\vec{\nabla}.\left(\vec{p}\delta^{\left(3\right)}\right)\label{eq:Main_RhoDef}\\
\vec{J}\left(t,\vec{r}\right)=\vec{\dot{p}}\delta^{\left(3\right)}+\vec{\nabla}\times\vec{\nabla}\times\left(c\vec{T}\delta^{\left(3\right)}\right)\label{eq:Main_JDef}\\
\vec{T}=\vec{N},\quad\vec{p}=\vec{\dot{N}}/c\label{eq:Main_AnaMomDef}
\end{gather}

Here, $c$ is the speed of light, $t$ and $\vec{r}$ are time and
the position of the observer, whilst $\vec{p}=\vec{p}\left(t\right)$
and $\vec{T}=\vec{T}\left(t\right)$ are the electric and toroidal
dipole moments (respectively). The three-dimensional delta function
is denoted with $\delta^{\left(3\right)}=\delta^{\left(3\right)}\left(\vec{r}-\vec{\bar{r}}\right)$,
where $\vec{\bar{r}}$ is the position of the point-particle. The
time-derivative is denoted by `overdot', i.e. $d\vec{p}/dt\equiv\vec{\dot{p}}$.
Apart from defining anapole excitation, Eqs.~(\ref{eq:Main_RhoDef},~\ref{eq:Main_JDef})
also provide the definition for the charge and current density of
a point-like electric dipole (obtained by setting $\vec{T}\to0$)
and a point-like toroidal dipole (obtained by setting $\vec{p}\to0$).
We now return to the definition of the anapole. To make an anapole,
the electric and toroidal dipoles in the equations above have to be
linked. A convenient way to do so is to introduce an `anapole dipole
moment' $\vec{N}$, defined by Eq.~(\ref{eq:Main_AnaMomDef}). 

Before proceeding, we introduce a convenient terminology. Clearly,
the anapole defined in Eqs.~(\ref{eq:Main_RhoDef}-\ref{eq:Main_AnaMomDef})
is just one possible kind of a nonradiating configuration. A suitable
super-position of electric and toroidal quadrupoles, for example,
can also lead to nonradiating configurations, which will differ from
the anapole introduced above. We therefore will refer to the excitations
defined in Eqs.~(\ref{eq:Main_RhoDef}-\ref{eq:Main_AnaMomDef})
as `anapole dipole' throughout this paper.

The definition provided in Eqs.~(\ref{eq:Main_RhoDef}-\ref{eq:Main_AnaMomDef})
is ill-suited for calculations involving moving and accelerating sources,
one therefore needs to generalize it to four-current of an electric/toroidal/anapole
dipole. This is accomplished in the next three sub-sections. Sub-sections~\ref{subsec:ElDipJ}
and~\ref{subsec:TDipJ} provide the correct expressions for the four-current
of a point-particle with electric dipole moment and toroidal dipole
moment (respectively). Superposition of these two four-currents, which
corresponds to an anapole, is given the Sub-section~\ref{subsec:AnaDipJ}.


\subsection{Four-current of a point-particle with electric dipole moment\label{subsec:ElDipJ}}

Here we consider the four-current ($J^{\left(p\right)\mu}$) due to
a point-particle with electric dipole moment ($\vec{p}$). That is,
we consider the four-current due to a point-particle in arbitrary
motion, which, when characterized in its \emph{instantaneous} rest
frame, has only the electric dipole moment $\vec{p}$. The four-current
can be derived in a number of ways, including simply considering a
pair of co-moving opposite charges that oscillate about a point $\bar{x}^{\mu}$,
which is the four-position of the electric dipole particle. Fortunately,
point-particles with electric dipole moment have already been considered
in the literature \citep{Ellis66}. The four-current due to electric
dipole point-particle is:
\begin{equation}
J^{\left(p\right)\mu}=\int d\tau\,c^{2}p^{\alpha}\hat{\tau}^{\beta}\,\delta_{\alpha\beta}^{\mu\eta}\nabla_{\eta}\delta^{\left(4\right)}\left(x-\bar{x}\right)\label{eq:Main_ElDipJ}
\end{equation}

Where $\tau$ is the proper time of the particle, $\bar{x}^{\mu}=\bar{x}^{\mu}\left(\tau\right)$
is the four-position of the particle, $\hat{\tau}^{\alpha}=\left(d\bar{x}^{\mu}/d\tau\right)/c$
is the unit-length four-vector tangent to the particle world-line,
$\nabla_{\eta}$ is the covariant derivative, $\delta_{\alpha\beta}^{\mu\eta}=\delta_{\alpha}^{\mu}\delta_{\beta}^{\eta}-\delta_{\beta}^{\mu}\delta_{\alpha}^{\eta}$
is the generalized Kroenecker delta \citep{LovelockBook}, and $p^{\alpha}$
is the four-vector related to the electric dipole moment of the particle.
One can demonstrate that Eq.~(\ref{eq:Main_ElDipJ}) is the correct
expression by considering its equivalent in the instantaneous rest
frame of the particle. Let frame $\tilde{S}$ be an inertial reference
frame in which the electric dipole particle is at rest at proper time
$\tau=\tau_{0}$ which corresponds to time $\tilde{t}=0$ in $\tilde{S}$.
Assume that velocity of the particle in $\tilde{S}$ remains negligible
over the time-period $\tilde{t}=-\frac{\epsilon}{2}\dots\frac{\epsilon}{2}$.
\textbf{Within this period} the four-current due to electric dipole
point-particle in $\tilde{S}$ is:
\begin{flalign*}
\tilde{J}^{\left(p\right)\mu} & =\int_{-\epsilon/2}^{\epsilon/2}\frac{d\tilde{t}'}{\gamma}\,c^{2}\tilde{p}^{\alpha}\tilde{\hat{\tau}}^{\beta}\,\tilde{\delta}_{\alpha\beta}^{\mu\eta}\tilde{\nabla}_{\eta}\Lambda\delta^{\left(4\right)}\left(\tilde{x}-\tilde{\bar{x}}\right)\\
\tilde{J}^{\left(p\right)\mu} & =\int_{-\epsilon/2}^{\epsilon/2}d\tilde{t}'\,c^{2}\tilde{p}^{\alpha}\,\tilde{\delta}_{\alpha0}^{\mu\eta}\tilde{\nabla}_{\eta}\delta^{\left(4\right)}\left(\tilde{x}-\tilde{\bar{x}}\right)
\end{flalign*}

Where $\tilde{\bar{x}}^{\mu}=\left(c\tilde{t},\,\tilde{\bar{\vec{r}}}\right)$
is the four-position of the particle, $\gamma=d\tilde{t}/d\tau=1/\sqrt{1-\dot{\bar{\vec{r}}}^{2}/c^{2}}$
is the Lorentz factor, $\Lambda=\det\left(\partial\left(\tilde{x}\right)/\partial\left(x\right)\right)$
is the Jacobian due to changing coordinates from lab-frame to $\tilde{S}$.
Since $\tilde{S}$ is the instantaneous rest-frame $\gamma=1$ and
$\tilde{\hat{\tau}}^{\beta}=\tilde{\delta}_{0}^{\beta}$ (i.e. temporal
axis of $\tilde{S}$ is tangent to particle world line at $\tilde{t}=0$).
We also assume that the the lab-frame and $\tilde{S}$ only differ
by a boost, thus $\Lambda=1$. 

Due to anti-symmetry of $\tilde{\delta}_{\alpha0}^{\mu\eta}$ the
four-vector $\tilde{p}^{\alpha}$ can be defined as space-like without
any loss of generality: $\tilde{p}^{\alpha}=\left(0,\,\vec{p}\right)^{\alpha}$.
The charge density then becomes:
\begin{flalign*}
\tilde{\rho} & =\tilde{J}^{\left(p\right)0}/c=\int_{-\epsilon/2}^{\epsilon/2}d\tilde{t}'\,c\tilde{p}^{\alpha}\,\tilde{\delta}_{\alpha0}^{0\eta}\tilde{\nabla}_{\eta}\delta\left(c\left(\tilde{t}-\tilde{t}'\right)\right)\delta^{\left(3\right)}\left(\vec{\tilde{r}}-\vec{\tilde{\bar{r}}}\right)\\
 & =-\int_{-\epsilon/2}^{\epsilon/2}d\tilde{t}'\,c\tilde{p}^{\alpha}\tilde{\nabla}_{\alpha}\delta\left(c\left(\tilde{t}-\tilde{t}'\right)\right)\delta^{\left(3\right)}\left(\vec{\tilde{r}}-\vec{\tilde{\bar{r}}}\right)\\
 & =-\vec{p}.\tilde{\vec{\nabla}}\delta^{\left(3\right)}\left(\vec{\tilde{r}}-\vec{\tilde{\bar{r}}}_{0}\right)=-\vec{p}.\tilde{\vec{\nabla}}\delta^{\left(3\right)}
\end{flalign*}

Which corresponds to charge density of a point-particle with electric
dipole moment $\vec{p}$ (Eq.~(\ref{eq:Main_RhoDef})). Above, $\vec{\tilde{\bar{r}}}_{0}$
denotes the position of the particle at time $\tilde{t}=0$ in frame
$\tilde{S}$. The current density is ($i=1,\,2,\,3$):
\begin{flalign*}
\tilde{\vec{J}}^{i} & =\int_{-\epsilon/2}^{\epsilon/2}d\tilde{t}'\,c^{2}\tilde{p}^{\alpha}\,\tilde{\delta}_{\alpha0}^{i\eta}\tilde{\nabla}_{\eta}\delta\left(c\left(\tilde{t}-\tilde{t}'\right)\right)\delta^{\left(3\right)}\left(\vec{\tilde{r}}-\vec{\tilde{\bar{r}}}\right)\\
 & =\int_{-\epsilon/2}^{\epsilon/2}d\tilde{t}'\,c^{2}\tilde{p}^{i}\,\tilde{\nabla}_{0}\delta\left(c\left(\tilde{t}-\tilde{t}'\right)\right)\delta^{\left(3\right)}\left(\vec{\tilde{r}}-\vec{\tilde{\bar{r}}}\right)\\
 & =\int_{-\epsilon/2}^{\epsilon/2}d\tilde{t}'\,c\tilde{p}^{i}\left(\tilde{t}'\right)\,\partial_{\tilde{t}}\delta\left(c\left(\tilde{t}-\tilde{t}'\right)\right)\delta^{\left(3\right)}\left(\vec{\tilde{r}}-\vec{\tilde{\bar{r}}}\right)\\
\tilde{\vec{J}} & =\left[\partial_{\tilde{t}}\left(\vec{p}\delta\left(\vec{\tilde{r}}-\vec{\tilde{\bar{r}}}\right)\right)\right]_{\tilde{t}=0}=\dot{\vec{p}}\delta^{\left(3\right)}-\vec{p}\left(\vec{\dot{\tilde{\bar{r}}}}_{0}.\tilde{\vec{\nabla}}\delta^{\left(3\right)}\right)=\dot{\vec{p}}\delta^{\left(3\right)}
\end{flalign*}

Note that by definition, the velocity of the point-particle is negligible
in $\tilde{S}$ during $\tilde{t}=-\epsilon/2\dots\epsilon/2$ so
$\vec{\dot{\tilde{\bar{r}}}}\left(0\right)=\vec{\dot{\tilde{\bar{r}}}}_{0}=0$.
The above expression agrees with the current density due to point-particle
with electric dipole $\vec{p}$ (Eq.~(\ref{eq:Main_JDef}) with $\vec{T}=0$).
It follows that Eq.~(\ref{eq:Main_ElDipJ}) is the four-current of
a point-particle which can be described as electric dipole point-particle
in its instantaneous rest frame ($\tilde{S}$) at any point in its
history. The four-vector $p^{\mu}$ is given by $p^{\mu}=\frac{\partial x^{\mu}}{\partial\tilde{x}^{\nu}}\left(0,\,\vec{p}\right)^{\nu}$,
where $\tilde{S}$ is the instantaneous rest-frame of the particle
and $\vec{p}$ is the electric dipole moment of the particle in $\tilde{S}$.

\subsection{Four-current of a point-particle with toroidal dipole moment\label{subsec:TDipJ}}

For toroidal dipole, we adopt an approach similar to Sec.~\ref{subsec:ElDipJ},
the expression for the four-current of a point-particle with toroidal
dipole moment is first stated:
\begin{equation}
J^{\left(T\right)\mu}=\int d\tau\,c^{2}T^{\gamma}\hat{\tau}^{\sigma}\hat{\tau}_{\eta}\delta_{\gamma\sigma\rho}^{\mu\eta\alpha}g^{\rho\beta}\nabla_{\alpha}\nabla_{\beta}\delta^{\left(4\right)}\left(x-\bar{x}\right)\label{eq:Main_TDipJ}
\end{equation}

and then it is shown that this expression does reduce to Eq.~(\ref{eq:Main_JDef})
(with $\vec{p}=\vec{0}$) in the instantaneous rest-frame of the particle.
Above $\hat{\tau}_{\eta}=g_{\eta\nu}\hat{\tau}^{\nu}$, where $g_{\eta\nu}=\text{diag}\left(1,-1,-1,-1\right)$
is the metric tensor and $g^{\rho\beta}$ is the inverse metric tensor.
The four-vector $T^{\gamma}$ is related to the toroidal dipole moment
of the point-particle.

As in Sec.~\ref{subsec:ElDipJ}, one goes into instantaneous rest
frame $\tilde{S}$ of the particle, and uses $d\tau=d\tilde{t}$,
$\tilde{\hat{\tau}}^{\sigma}=\tilde{\delta}_{0}^{\sigma}$, $\tilde{\hat{\tau}}_{\eta}=\tilde{\delta}_{\eta}^{0}$
:
\begin{flalign*}
\tilde{J}^{\left(T\right)\mu} & =\int_{-\epsilon/2}^{\epsilon/2}d\tilde{t}'\,c^{2}\tilde{T}^{\gamma}\tilde{\delta}_{\gamma0\rho}^{\mu0\alpha}\tilde{g}^{\rho\beta}\tilde{\nabla}_{\alpha}\tilde{\nabla}_{\beta}\delta^{\left(4\right)}\left(\tilde{x}-\tilde{\bar{x}}\right)\\
\tilde{\rho} & =\tilde{J}^{\left(T\right)0}/c=0\\
\tilde{\vec{J}}^{\left(T\right)i} & =c\left(0,\vec{T}\right)^{\gamma}\tilde{\delta}_{\gamma0\rho}^{i0\alpha}\tilde{g}^{\rho\beta}\tilde{\nabla}_{\alpha}\tilde{\nabla}_{\beta}\delta^{\left(3\right)}\left(\vec{\tilde{r}}-\vec{\tilde{\bar{r}}}_{0}\right)\\
 & =c\left(\begin{array}{c}
-T_{x}\tilde{\vec{\nabla}}^{2}+\left(\vec{T}.\tilde{\vec{\nabla}}\right)\tilde{\partial}_{x}\\
-T_{y}\tilde{\vec{\nabla}}^{2}+\left(\vec{T}.\tilde{\vec{\nabla}}\right)\tilde{\partial}_{y}\\
-T_{z}\tilde{\vec{\nabla}}^{2}+\left(\vec{T}.\tilde{\vec{\nabla}}\right)\tilde{\partial}_{z}
\end{array}\right)\delta^{\left(3\right)}\left(\vec{\tilde{r}}-\vec{\tilde{\bar{r}}}_{0}\right)\\
 & =\tilde{\vec{\nabla}}\times\tilde{\vec{\nabla}}\times c\vec{T}\delta^{\left(3\right)}
\end{flalign*}

Above, the anti-symmetry of $\delta_{\gamma\sigma\rho}^{\mu\eta\alpha}=\delta_{\gamma}^{\mu}\delta_{\sigma\rho}^{\eta\alpha}-\delta_{\sigma}^{\mu}\delta_{\gamma\rho}^{\eta\alpha}-\delta_{\rho}^{\mu}\delta_{\sigma\gamma}^{\eta\alpha}$
ensures that $\tilde{J}^{\left(T\right)0}=0$ and that $\tilde{T}^{0}$
has no effect on the four-current, hence one can set $\tilde{T}^{0}=0$.
Thus the four-current in Eq.~(\ref{eq:Main_TDipJ}) corresponds to
point-particle which has toroidal dipole moment $\vec{T}$ in its
instantaneous rest frame at all times (compare with Eq.~(\ref{eq:Main_JDef})
when $\vec{p}=0$). As with electric dipole, the four-vector $T^{\gamma}=\frac{\partial x^{\mu}}{\partial\tilde{x}^{\nu}}\left(0,\,\vec{T}\right)^{\nu}$
where $\tilde{S}$ is the instantaneous rest frame and $\vec{T}$
is the toroidal dipole moment of the point-particle in $\tilde{S}$.

\subsection{Four-current of a point-particle with anapole moment\label{subsec:AnaDipJ}}

Combining the previous results in Eq.~(\ref{eq:Main_ElDipJ},\ref{eq:Main_TDipJ})
the four-current of the anapole point-particle is:
\begin{multline}
J^{\mu}\left(x^{\nu}\right)=\int d\tau\,\left(c\frac{dN^{\gamma}}{d\tau}\hat{\tau}^{\sigma}\delta_{\gamma\sigma}^{\mu\alpha}\;\nabla_{\alpha}\delta^{\left(4\right)}+\right.\\
\left.+c^{2}N^{\gamma}\hat{\tau}^{\sigma}\hat{\tau}_{\eta}\delta_{\gamma\sigma\rho}^{\mu\eta\alpha}g^{\rho\beta}\nabla_{\alpha}\nabla_{\beta}\delta^{\left(4\right)}\right)\label{eq:Main_AnaFourCurrent}
\end{multline}

The four-vector $N^{\gamma}$ is such that in the \emph{instantaneous}
rest-frame of the particle ($\tilde{S}$) it has value $\tilde{N}^{\gamma}=\left(0,\,\vec{N}\right)^{\gamma}$,
where $\vec{N}$ is the anapole moment of Eq.~(\ref{eq:Main_AnaMomDef}). 

To the best of our knowledge, this is the first time the covariant
description of either the toroidal dipole or anapole has been presented. 

\section{Radiation from moving point-particle multipoles\label{sec:Radiation}}

In this section we will show how the radiation field due to arbitrary
point-dipoles can be evaluated in a straight-forward fashion. As shown
by several authors \citep{Ellis63,Ward65,Ellis66}, four-currents
due to moving point-dipoles can be expressed as linear combinations
of derivatives of delta-functions integrated along the world-line
of the point-particle: 
\begin{equation}
J^{\mu}\left(x\right)=\int d\tau\,K^{\mu\alpha_{1}\dots\alpha_{n}}\nabla_{\alpha_{1}}\dots\nabla_{\alpha_{n}}\delta^{\left(4\right)}\left(x-\bar{x}\left(\tau\right)\right)\label{eq:Main_GenericPointDipoleJ}
\end{equation}

Where $K^{\mu\alpha_{1}\dots\alpha_{n}}$ is a tensor that depends
on proper time $\tau$ and $\bar{x}^{\mu}\left(\tau\right)$ is the
world-line of the point-particle. Indeed, the four-currents for electric
(Eq.~(\ref{eq:Main_ElDipJ})) and toroidal (Eq.~(\ref{eq:Main_TDipJ}))
point-dipoles can be expressed in the form of Eq.~(\ref{eq:Main_GenericPointDipoleJ}).
The four-potential ($A^{\mu}=\left(\phi/c,\,\vec{A}\right)^{\mu}$)
due to four-current in Eq.~(\ref{eq:Main_GenericPointDipoleJ}) is
given by $\partial_{\mu}\partial^{\mu}A^{\eta}=\mu_{0}J^{\eta}$,
where $\mu_{0}$ is vacuum permeability \citep{jackson}. Substituting
the above current density one can find the retarded solution for the
four-potential \citep{jackson}, and from that the electromagnetic
tensor:
\begin{flalign}
F_{\mu\eta}= & \delta_{\mu\eta}^{\gamma\xi}\nabla_{\gamma}A_{\xi}\nonumber \\
= & \frac{\mu_{0}}{2\pi}\,\delta_{\mu\eta}^{\gamma\xi}\,g_{\xi\kappa}\nabla_{\gamma}\nabla_{\alpha_{1}}\dots\nabla_{\alpha_{n}}\biggl[\nonumber \\
 & \int d\tau\,K^{\kappa\alpha_{1}\dots\alpha_{n}}\Theta\left(X^{0}\right)\delta\left(X\cdot X\right)\biggr]\label{eq:Main_EMTensGeneral}
\end{flalign}

Where $X^{\eta}=x^{\eta}-\bar{x}^{\eta}$ is a four-vector connecting
the four-position of the observer at $x^{\eta}=\left(ct,\,\vec{r}\right)^{\eta}$
and the four-position of the point-particle at $\bar{x}^{\eta}=\left(c\bar{t},\,\bar{\vec{r}}\right)^{\eta}$.
The Heaviside function $\Theta\left(X^{0}\right)=\Theta\left(c\left(t-\bar{t}\right)\right)$
enforces causality. We will bear this constraint in mind and omit
$\Theta\left(X^{0}\right)$ in what is to follow (i.e. we will restrict
our consideration to $t>\bar{t}$, which implies an upper limit on
the proper time integral $\int d\tau\to\int^{\tau\left(t\right)}d\tau$
,which we will also omit for clarity). 

In principle Eq.~(\ref{eq:Main_EMTensGeneral}) can be evaluated
in full generality, but the resulting expression tend to become very
cumbersome very fast. Instead we offer a short-cut which is capable
of pulling out only the radiation fields, i.e. fields that decay as
$1/\left|\vec{r}-\bar{\vec{r}}\right|$ with distance between the
source and the observer. It should be noted, that near-field of point-like
emitters can contain a wealth of interesting information on emission
and re-absorption of electromagnetic energy \citep{Mandel72,Bossy12},
however here we shall forgo this in the interest of simplicity. 

The short-cut can be illustrated by the simple case of $n=0$ and
$K^{\mu}=qc^{2}\hat{\tau}^{\mu}$ (current density due to point-charge
\citep{jackson}):
\begin{flalign*}
F_{\mu\eta}^{(n=0)}= & \frac{\mu_{0}}{2\pi}\,\delta_{\mu\eta}^{\gamma\xi}\,g_{\xi\kappa}\int d\tau\,K^{\kappa}\nabla_{\gamma}\delta\left(X\cdot X\right)
\end{flalign*}

Following Ref.~\citep{jackson} we rewrite ($u^{\mu}=d\bar{x}^{\mu}/d\tau$
is the four-velocity):
\begin{flalign}
\nabla_{\gamma}\delta\left(X\cdot X\right)= & \nabla_{\gamma}\delta\left(\left(x-\bar{x}\right)_{\sigma}\left(x-\bar{x}\right)^{\sigma}\right)\nonumber \\
= & -\frac{X_{\gamma}}{u\cdot X}\,\,\frac{d}{d\tau}\delta\left(X\cdot X\right)\label{eq:Main_XAppers}
\end{flalign}

Substituting and integrating by parts:
\begin{flalign}
F_{\mu\eta}^{(n=0)}= & \frac{\mu_{0}}{2\pi}\,\delta_{\mu\eta}^{\gamma\xi}\,g_{\xi\kappa}\int d\tau\,\left(-\frac{X_{\gamma}}{u\cdot X}\right)K^{\kappa}\,\frac{d}{d\tau}\delta\left(X\cdot X\right)\nonumber \\
= & \frac{\mu_{0}}{2\pi}\,\delta_{\mu\eta}^{\gamma\xi}\,g_{\xi\kappa}\int d\tau\,\frac{d}{d\tau}\left[\left(\frac{X_{\gamma}}{u\cdot X}\right)K^{\kappa}\right]\,\delta\left(X\cdot X\right)\label{eq:Main_EMtensN0}
\end{flalign}

The one-dimensional delta-function can be re-cast into \citep{jackson}:

\begin{gather*}
\delta\left(X\cdot X\right)=\frac{1}{2\left(u\cdot X\right)}\delta\left(\tau-\tau_{ret}\right)
\end{gather*}

Where $\tau_{ret}$ is the proper time at which $X^{\sigma}=\left(r,\,\vec{r}\right)^{\sigma}$,
i.e. when the observer located at position $\vec{r}$ away from the
point-particle observes the radiation emitted by the point-particle
at time $r/c$ earlier. For any function of proper time $f=f\left(\tau\right)$
\citep{jackson}:

\begin{flalign*}
\int d\tau\,\delta\left(X\cdot X\right)f\left(\tau\right)= & \left[\frac{f\left(\tau\right)}{2\left(u\cdot X\right)}\right]_{\tau=\tau_{ret}}=\frac{f\left(\tau_{ret}\right)}{2r\gamma\left(c-\vec{\hat{r}}.\vec{v}\right)}
\end{flalign*}

Above $\vec{v}=d\bar{\vec{r}}/dt$ is the particle velocity in frame
$S$ (at time that corresponds to $\tau_{ret}$), and $\gamma=1/\sqrt{1-\left(v/c\right)^{2}}$
is the Lorentz factor. The key observation about the above equation
is that it is already decaying as $1/r$ with the distance between
the point-multipole and the observer. Therefore if one wants the integral
above to decay as $1/r$, $f\left(\tau_{ret}\right)$ must not decay
with $r$ at all, which means that it has to be zeroth order in $X^{\sigma}$
(since $X^{\sigma}\to r\left(1,\,\vec{\hat{r}}\right)^{\sigma}$ when
$\tau\to\tau_{ret}$).

Returning to Eq.~(\ref{eq:Main_EMtensN0}), and using $u^{\mu}=d\bar{x}/d\tau$
and $a^{\mu}=du^{\mu}/d\tau$:

\begin{multline*}
\frac{d}{d\tau}\left[\left(\frac{X_{\gamma}}{u\cdot X}\right)K^{\kappa}\right]=\frac{-u_{\gamma}}{u\cdot X}K^{\kappa}-\\
-\frac{X_{\gamma}}{\left(u\cdot X\right)^{2}}K^{\kappa}\left(a\cdot X-c^{2}\right)+\frac{X_{\gamma}}{u\cdot X}\,\frac{d}{d\tau}\left(K^{\kappa}\right)
\end{multline*}

Above, one can see that all the terms obtained by differentiating
$X$ will be of order $1/X$ and will therefore not contribute to
far-field radiation. One can also see that this has to be a general
prescription. Function $K^{\dots}$ does not depend on $X$, and the
only way that $X$ appears under the integral is via Eq.~(\ref{eq:Main_XAppers}).
However, at that point the whole expression is zeroth order in $X$
($X$ appears in the numerator and the denominator), thus any differentiation
of $X$ will make the expression negative order in $X$. \textbf{Therefore
the rule for obtaining only the far-field radiation contributions
becomes: not to differentiate $X$.} We shall denote this by putting
a dot above the equals sign, so, for example:
\begin{multline*}
\frac{d}{d\tau}\left[\left(\frac{X_{\gamma}}{u\cdot X}\right)K^{\kappa}\right]\doteq-\frac{X_{\gamma}\left(a\cdot X\right)}{\left(u\cdot X\right)^{2}}K^{\kappa}+\frac{X_{\gamma}}{u\cdot X}\,\frac{d}{d\tau}\left(K^{\kappa}\right)
\end{multline*}

This rule allows to recast Eq.~(\ref{eq:Main_EMTensGeneral}) into
a form more suitable for evaluation:

\begin{widetext}
\begin{flalign}
F_{\mu\eta}\doteq & \frac{\mu_{0}}{2\pi}\,\delta_{\mu\eta}^{\gamma\xi}\,g_{\xi\kappa}\int d\tau\,\frac{d}{d\tau}\left[\left(\frac{X_{\gamma}}{u\cdot X}\right)\frac{d}{d\tau}\left[\left(\frac{X_{\alpha_{1}}}{u\cdot X}\right)\dots\frac{d}{d\tau}\left[\left(\frac{X_{\alpha_{n}}}{u\cdot X}\right)K^{\kappa\alpha_{1}\dots\alpha_{n}}\right]\dots\right]\right]\delta\left(X\cdot X\right)\label{eq:Main_EMTensorFromK}\\
F_{\mu\eta}\doteq & \frac{\mu_{0}}{4\pi}\,\delta_{\mu\eta}^{\gamma\xi}\,g_{\xi\kappa}\left[\frac{1}{\left(u\cdot X\right)}\frac{d}{d\tau}\left[\left(\frac{X_{\gamma}}{u\cdot X}\right)\frac{d}{d\tau}\left[\left(\frac{X_{\alpha_{1}}}{u\cdot X}\right)\dots\frac{d}{d\tau}\left[\left(\frac{X_{\alpha_{n}}}{u\cdot X}\right)K^{\kappa\alpha_{1}\dots\alpha_{n}}\right]\dots\right]\right]\right]_{\tau=\tau_{ret}}\label{eq:Main_EMTensorFromKEval}
\end{flalign}

\end{widetext}

\section{Radiation from uniformly accelerating point-particle electric, toroidal
and anapole dipoles\label{sec:RadiationCalc}}

In this section we will apply Eq.~(\ref{eq:Main_EMTensorFromKEval})
to find the far-field radiation from point-particle with electric
(Sec.~\ref{subsec:UniAccEDip}), toroidal (Sec.~\ref{subsec:UniAccTDip}),
and anapole (Sec.~\ref{subsec:UniAccAna}) dipolar moments.

Before proceeding we shall briefly review the expressions specific
to motion of uniformly accelerated point-particles. The world-line
of a point-particle, in the lab-frame, with uniform acceleration $a$,
along the z-axis, is \citep{MTWBook}:

\begin{flalign}
\bar{x}^{\mu}\left(\tau\right)= & \frac{c^{2}}{a}\left(\sinh\frac{a\tau}{c},0,0,\cosh\frac{a\tau}{c}-1\right)^{\mu}\label{eq:Main_ParticleWorldlineTau}\\
\bar{x}^{\mu}\left(t\right)= & \left(ct,0,0,\frac{c^{2}}{a}\left(\sqrt{1+\left(at/c\right)^{2}}-1\right)\right)^{\mu}\label{eq:Main_ParticleWorldlineTee}
\end{flalign}

Above, $\tau$ is the proper time of the accelerating particle and
the over-bar above $\bar{x}^{\mu}$ is introduced used to distinguish
the position of the accelerating particle from the location of a generic
event. The integration constants are chosen so that at time $t=0$,
which also corresponds to proper time $\tau=0$, the accelerating
particle is passing through the lab-frame origin, and is instantaneously
at rest in the lab-frame. The parameter $a=\sqrt{-a^{\mu}a_{\mu}}$
is the magnitude of the four-acceleration $a^{\mu}=d^{2}\bar{x}^{\mu}/d\tau^{2}$. 

\begin{figure}
\begin{centering}
\includegraphics{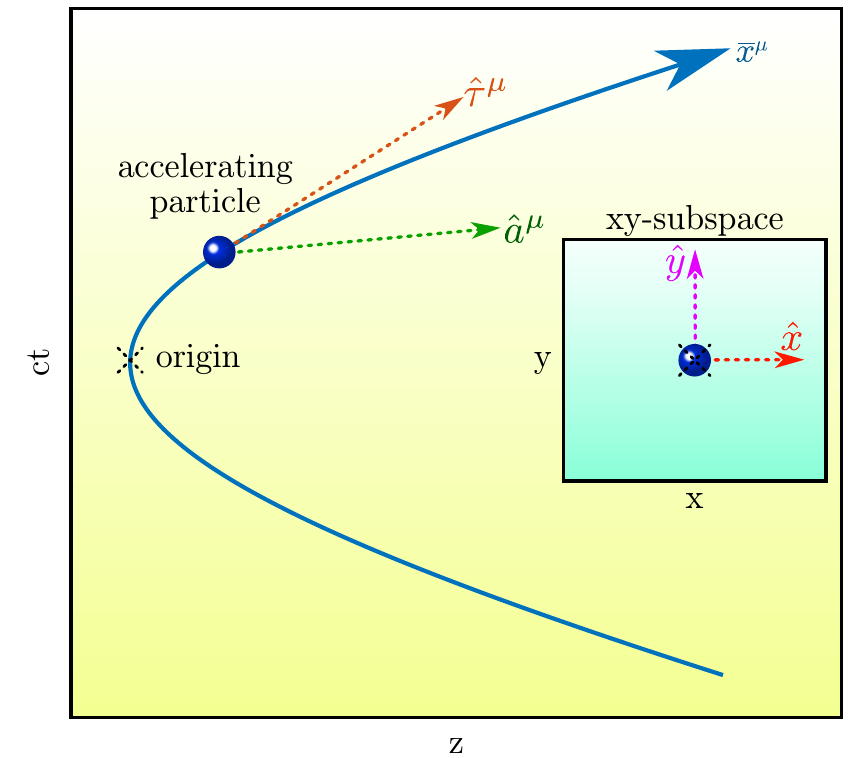}
\par\end{centering}
\caption{\textbf{Worldline of the uniformly accelerating particle.} Blue curve
(hyperbola) shows the path of the uniformly accelerating particle
(blue dot). Vectors $\hat{\tau}^{\mu}$ and $\hat{a}^{\mu}$ are orthonormal
vectors in the direction of instantaneous four-velocity, and four-acceleration.
Together they span $tz$-subspace. The inset shows two other orthonormal
vector $\hat{x}^{\mu}$ and $\hat{y}^{\mu}$ which span the $xy$-subspace.
The four vectors form a complete orthonormal set that spans the space-time.
Speed of light is denoted by $c$.}

\label{fig:Main_BasisDef}
\end{figure}

The worldline of the observer (following Eq.~(\ref{eq:Main_ParticleWorldlineTau}))
is shown in Fig.~\ref{fig:Main_BasisDef}. It is convenient to introduce
four four-vectors that form an orthonormal set that spans the spacetime.
Vector $\hat{\tau}^{\mu}$ points in the direction tangent to the
worldline of the accelerating particle, and is therefore parallel
to four-velocity $d\bar{x}^{\mu}/d\tau=u^{\mu}=c\hat{\tau}^{\mu}$
($\hat{\tau}\cdot\hat{\tau}=\hat{\tau}^{\mu}\hat{\tau}_{\mu}=1$),
vector $\hat{a}^{\mu}$ is parallel to four-acceleration $a^{\mu}=du^{\mu}/d\tau=a\hat{a}^{\mu}$
($\hat{a}\cdot\hat{a}=-1$), finally vectors $\hat{x}^{\mu}$ and
$\hat{y}^{\mu}$ are along the $x$- and $y$-axis ($\hat{x}\cdot\hat{x}=-1$
and $\hat{y}\cdot\hat{y}=-1$).

From Eq.~(\ref{eq:Main_ParticleWorldlineTau}) it follows:

\begin{flalign*}
\frac{du^{\mu}}{d\tau}= & a^{\mu}=a\hat{a}^{\mu} & \frac{d\hat{a}^{\mu}}{d\tau}= & \frac{a}{c}\hat{\tau}^{\mu}\\
\frac{d\hat{\tau}^{\mu}}{d\tau}= & \frac{a}{c}\hat{a}^{\mu} & \frac{da^{\mu}}{d\tau}= & \left(\frac{a}{c}\right)^{2}u^{\mu}=\frac{a^{2}}{c}\hat{\tau}^{\mu}
\end{flalign*}

Both in case of electric (Eq.~(\ref{eq:Main_ElDipJ})) and toroidal
(Eq.~(\ref{eq:Main_TDipJ})) dipole we will be working with vectors
that are, by definition, always orthogonal to the world-line of the
particle, i.e. $p\cdot\hat{\tau}=p^{\mu}\hat{\tau}_{\mu}=0$ for electric
dipole four-vector (and the same for $T^{\mu}$). Such vectors can
be conveniently parametrized by:
\[
p^{\mu}=p^{1}\hat{x}^{\mu}+p^{2}\hat{y}^{\mu}+p^{a}\hat{a}^{\mu}
\]

We shall be interested in proper-time derivatives of such vectors.
Introducing $\dot{p}^{\mu}\equiv\frac{dp^{1}}{d\tau}\hat{x}^{\sigma}+\frac{dp^{2}}{d\tau}\hat{y}^{\sigma}+\frac{dp^{a}}{d\tau}\hat{a}^{\sigma}$
(and the corresponding equivalent for $\ddot{p}^{\sigma}$ etc.),
one finds:
\begin{flalign*}
\frac{dp^{\mu}}{d\tau}= & \dot{p}^{\mu}+p^{a}\frac{d\hat{a}^{\mu}}{d\tau}=\dot{p}^{\mu}+\left(-p\cdot\hat{a}\right)\left(\frac{a}{c}\hat{\tau}^{\mu}\right)\\
= & \dot{p}^{\mu}-\frac{a}{c}\left(p\cdot\hat{a}\right)\hat{\tau}^{\mu}
\end{flalign*}

\subsection{Uniformly accelerated electric dipole point-particle\label{subsec:UniAccEDip}}

The suitable current density for electric point-particle with electric
dipole is given in Eq.~(\ref{eq:Main_ElDipJ}). Substituting it into
Eq.~(\ref{eq:Main_EMTensorFromKEval}), the electromagnetic tensor
for the far-field radiation from a point-particle electric dipole
becomes:
\begin{multline*}
F_{\mu\eta}^{(p)}\doteq\frac{\mu_{0}c^{2}}{2\pi}\delta_{\mu\eta}^{\gamma\xi}\delta_{\sigma\rho}^{\kappa\alpha}g_{\xi\kappa}\left[\frac{1}{2\left(u\cdot X\right)}\,\frac{d}{d\tau}\biggl[\right.\\
\left.\left.\left(\frac{X_{\gamma}}{u\cdot X}\right)\frac{d}{d\tau}\left[\left(\frac{X_{\alpha}}{u\cdot X}\right)p^{\sigma}\hat{\tau}^{\rho}\right]\right]\right]_{\tau=\tau_{ret}}
\end{multline*}

next one applies the relations from the introductory part of Sec.~\ref{sec:RadiationCalc}
to find radiation in the case of uniform acceleration:
\begin{multline}
F_{\mu\eta}^{(p)}\doteq\frac{\mu_{0}c^{2}}{4\pi}\delta_{\mu\eta}^{\gamma\xi}\delta_{\sigma\rho}^{\kappa\alpha}g_{\xi\kappa}\left[\frac{X_{\alpha}X_{\gamma}}{\left(u\cdot X\right)^{3}}\,\Biggl[\right.\\
\left(\ddot{p}^{\sigma}-3\left(\frac{a\cdot X}{u\cdot X}\right)\dot{p}^{\sigma}+3\left(\frac{a\cdot X}{u\cdot X}\right)^{2}p^{\sigma}\right)\hat{\tau}^{\rho}+\\
+\frac{a}{c}\left(2\dot{p}^{\sigma}-3\left(\frac{a\cdot X}{u\cdot X}\right)p^{\sigma}\right)\hat{a}^{\rho}-\\
\left.\left.-\left(\frac{a}{c}\right)^{2}\left(p\cdot\hat{a}\right)\hat{\tau}^{\sigma}\hat{a}^{\rho}\right]\right]_{\tau=\tau_{ret}}\label{eq:Main_EDipFieldDerivEvaluated}
\end{multline}

Above, the anti-symmetric property of Kronecker delta was used implicitly,
e.g. $\delta_{\sigma\rho}^{\kappa\alpha}\hat{\tau}^{\sigma}\hat{\tau}^{\rho}=0$.
Next, we simplify the expression by assuming that $a\tau/c\to0$,
i.e. we will assume that during the observation period the speed of
the accelerating point-particle is negligible in the lab-frame. At
the same time the particle will be located at the origin in the lab-frame
(see Eq.~(\ref{eq:Main_ParticleWorldlineTau}) and Fig.~\ref{fig:Main_BasisDef}).
One important consequence of this approximation is that electric dipole
of the particle in the lab-frame becomes the same as in its (instantaneous)
rest-frame, one can therefore talk about a three-dimensional dipole
moment vector $\vec{p}$ without ambiguity. Some of the transformations
due to this approximation are (see Eq.~(\ref{eq:Main_ParticleWorldlineTau})
and Fig.~\ref{fig:Main_BasisDef}):
\begin{gather}
\begin{array}{ccccc}
\hat{a}^{\mu}\to\hat{z}^{\mu}=\delta_{3}^{\mu}, & \: & p^{\mu}\to\left(0,\,\vec{p}\right)^{\mu}, & \: & \hat{\tau}^{\mu}\to\hat{t}^{\mu}=\delta_{0}^{\mu}\\
u^{\mu}\rightarrow c\delta_{0}^{\mu}, & \: & \tau_{ret}\rightarrow t-r/c, & \: & \left(u\cdot X\right)\rightarrow rc\\
\left(a\cdot X\right)\rightarrow-ar\,\vec{\hat{r}}.\vec{\hat{a}} & \: & X^{\mu}\to r\left(1,\,\vec{\hat{r}}\right)^{\mu}, & \: & \left(\hat{a}\cdot p\right)\to-\vec{\hat{a}}.\vec{p}
\end{array}\label{eq:Main_LabFrameApprox}
\end{gather}

It is important to note that this approximation can only be applied
at the end, once all derivatives have been evaluated. This last approximation
will be denoted by asterix over the equals sign ($\asteq$). Next,
we evaluate Eq.~(\ref{eq:Main_EDipFieldDerivEvaluated}) and use
the result to find electric ($\left(\vec{E}^{(p)}\right)^{i}=-cg^{ij}F_{0j}^{(p)}$)
and magnetic fields ($\left(\mu_{0}\vec{H}^{\left(p\right)}\right)^{i}=-\epsilon^{iab}F_{ab}^{\left(p\right)}/2$):

\begin{flalign}
\vec{E}^{(p)}\asteq & \frac{\mu_{0}}{4\pi r}\,\left[\vec{\hat{r}}\times\vec{\hat{r}}\times\boldsymbol{\mathcal{V}}-\right.\nonumber \\
 & \left.-\left(\frac{a}{c}\right)\vec{\hat{r}}\times\vec{\hat{a}}\times\boldsymbol{\mathcal{W}}-\left(\frac{a}{c}\right)^{2}\left(\vec{p}.\vec{\hat{a}}\right)\vec{\hat{r}}\times\vec{\hat{r}}\times\vec{\hat{a}}\right]\label{eq:Main_EDipElecF}\\
\vec{H}^{(p)}\asteq & \frac{1}{\mu_{0}c}\vec{\hat{r}}\times\vec{E}^{(p)}\label{eq:Main_EDipMagF}\\
\boldsymbol{\mathcal{V}}= & \ddot{\vec{p}}+3\frac{a}{c}\left(\vec{\hat{a}}.\vec{\hat{r}}\right)\dot{\vec{p}}+3\left(\frac{a}{c}\right)^{2}\left(\vec{\hat{a}}.\vec{\hat{r}}\right)^{2}\vec{p}\nonumber \\
\boldsymbol{\mathcal{W}}= & 2\dot{\vec{p}}+3\frac{a}{c}\left(\vec{\hat{a}}.\vec{\hat{r}}\right)\vec{p}\nonumber \\
\vec{p}= & \vec{p}\left(t-r/c\right)\nonumber 
\end{flalign}

Above we have introduced vector fields $\vec{\mathcal{V}}$ and $\vec{\mathcal{W}}$
for convenience. Here and for the rest of the text we remove the nested
brackets from cross-products, but our convention is to apply them
from right to left, i.e. $\vec{U}\times\vec{V}\times\vec{W}=\vec{U}\times\left(\vec{V}\times\vec{W}\right)$.
Both electric and magnetic fields are manifestly transverse, since
they can be written as $\vec{\hat{r}}\times...$. It follows that
$\vec{E}^{(p)}\times\vec{H}^{\left(p\right)}=\frac{1}{\mu_{0}c}\left(\vec{\hat{r}}\left|\vec{E}^{\left(p\right)}\right|^{2}-\vec{E}^{\left(p\right)}\left(\vec{\hat{r}}.\vec{E}^{\left(p\right)}\right)\right)=\left(\left|\vec{E}^{\left(p\right)}\right|^{2}/\mu_{0}c\right)\vec{\hat{r}}$.
Thus, as one should expect, the fields result in Poynting vector in
positive radial direction, i.e. electromagnetic energy is propagating
away from the dipole.

\subsection{Uniformly accelerated toroidal dipole point-particle\label{subsec:UniAccTDip}}

The suitable current density is given in Eq.~(\ref{eq:Main_TDipJ}).
Substituting it into Eq.~(\ref{eq:Main_EMTensorFromKEval}) results
in the following electromagnetic tensor:

\begin{widetext}
\begin{flalign}
F_{\mu\eta}^{\left(T\right)}\doteq & \frac{\mu_{0}c^{2}}{4\pi}\delta_{\mu\eta}^{\theta\xi}\,g_{\xi\kappa}\delta_{\phi\sigma\rho}^{\kappa\nu\alpha}g^{\rho\beta}\left[\frac{1}{u\cdot X}\,\frac{d}{d\tau}\left[\left(\frac{X_{\theta}}{u\cdot X}\right)\frac{d}{d\tau}\left[\left(\frac{X_{\alpha}}{u\cdot X}\right)\frac{d}{d\tau}\left[\left(\frac{X_{\beta}}{u\cdot X}\right)T^{\phi}\hat{\tau}^{\sigma}\hat{\tau}_{\nu}\right]\right]\right]\right]_{\tau=\tau_{ret}}\label{eq:Main_FToro}
\end{flalign}

\end{widetext}

Evaluation of the above expression is significantly more difficult
than Eq.~(\ref{eq:Main_EDipFieldDerivEvaluated}): there a more derivatives
and more terms. The details of the evaluation are therefore given
in the App.~\ref{app_subsec:toroRad}, whilst here we give only the
final result:
\begin{flalign}
\vec{E}^{\left(T\right)}\asteq & \frac{\mu_{0}}{4\pi rc}\left[-\vec{\hat{r}}\times\vec{\hat{r}}\times\vec{\mathcal{F}}+3\left(\frac{a}{c}\right)\vec{\hat{r}}\times\vec{\hat{a}}\times\vec{\mathcal{Q}}+\right.\nonumber \\
 & \left.\quad+3\left(\frac{a}{c}\right)^{2}\left(\vec{\hat{a}}.\vec{\mathcal{G}}\right)\vec{\hat{r}}\times\vec{\hat{r}}\times\vec{\hat{a}}\right]\label{eq:Main_TDipElecF}\\
\vec{H}^{\left(T\right)}\asteq & \frac{1}{\mu_{0}c}\vec{\hat{r}}\times\vec{E}^{\left(T\right)}\label{eq:Main_TDipMagF}
\end{flalign}

With:
\begin{flalign*}
\vec{\mathcal{F}}= & \dddot{\vec{T}}+3\left(\frac{a}{c}\right)\left(\vec{\hat{a}}.\vec{\hat{r}}\right)\ddot{\vec{T}}+\left(\frac{a}{c}\right)^{2}\left(2+3\left(\vec{\hat{a}}.\vec{\hat{r}}\right)^{2}\right)\dot{\vec{T}}+\\
 & \quad+3\left(\frac{a}{c}\right)^{3}\left(\vec{\hat{a}}.\vec{\hat{r}}\right)\vec{T}\\
\vec{\mathcal{Q}}= & \ddot{\vec{T}}+2\left(\frac{a}{c}\right)\left(\vec{\hat{a}}.\vec{\hat{r}}\right)\dot{\vec{T}}+\left(\frac{a}{c}\right)^{2}\left(\vec{\hat{a}}.\vec{\hat{r}}\right)^{2}\vec{T}\\
\vec{\mathcal{G}}= & \dot{\vec{T}}+\left(\frac{a}{c}\right)\left(\vec{\hat{a}}.\vec{\hat{r}}\right)\vec{T}\\
\vec{T}= & \vec{T}\left(t-r/c\right)
\end{flalign*}

Where $\vec{T}$ is the toroidal dipole moment of the point-particle.

\subsection{Uniformly accelerated anapole point-particle\label{subsec:UniAccAna}}

Due to linearity of Maxwell's equations the radiation pattern of the
accelerated point-particle with anapole moment can be obtained by
combining Eqs.~(\ref{eq:Main_EDipElecF},~\ref{eq:Main_EDipMagF},~\ref{eq:Main_TDipElecF},~\ref{eq:Main_TDipMagF})
with Eq.~(\ref{eq:Main_AnaMomDef}). After some simplification one
can show that the far-field, zero-velocity radiation of a point-particle
with anapole moment is:
\begin{flalign}
\vec{E}^{\left(N\right)}\asteq & \frac{\mu_{0}}{4\pi rc}\left(\frac{a}{c}\right)\left[\vec{\hat{r}}\times\vec{\hat{a}}\times\vec{\vec{\mathcal{R}}}+\right.\nonumber \\
 & \left.\quad\quad\quad\quad+\left(\frac{a}{c}\right)\vec{\hat{r}}\times\vec{\hat{r}}\times\vec{\hat{a}}\times\vec{\hat{a}}\times\vec{\mathcal{S}}\right]\label{eq:Main_AnaElecF}\\
\vec{H}^{\left(N\right)}\asteq & \frac{1}{\mu_{0}c}\vec{\hat{r}}\times\vec{E}^{\left(N\right)}\label{eq:Main_AnaMagF}
\end{flalign}

With:
\begin{flalign*}
\vec{\mathcal{R}}= & \ddot{\vec{N}}+3\left(\frac{a}{c}\right)\left(\vec{\hat{a}}.\vec{\hat{r}}\right)\dot{\vec{N}}+3\left(\frac{a}{c}\right)^{2}\left(\vec{\hat{a}}.\vec{\hat{r}}\right)^{2}\vec{N}\\
\vec{\mathcal{S}}= & 2\dot{\vec{N}}+3\left(\frac{a}{c}\right)\left(\vec{\hat{a}}.\vec{\hat{r}}\right)\vec{N}\\
\vec{N}= & \vec{N}\left(t-r/c\right)
\end{flalign*}

Where $\vec{N}$ is the anapole dipole moment as defined in Eq.~(\ref{eq:Main_AnaMomDef}).
We note in passing, that whilst in general the anapole particle does
not appear as `pure' anapole to a stationary observer, under zero-velocity
approximation, the current density of anapole particle, in Eq.~(\ref{eq:Main_AnaFourCurrent}),
becomes equivalent to `conventional' anapole, in Eq.~(\ref{eq:Main_RhoDef},\ref{eq:Main_JDef},\ref{eq:Main_AnaMomDef}). 

\section{Discussion\label{sec:Discussion}}

In this section we shall develop several properties of the light emitted
by accelerated electric, toroidal and anapole dipoles. The radiation
patterns of the emitted radiation are discussed in Sec.~\ref{subsec:RadPat}.
Section~\ref{subsec:Ellipticity} discusses how the ellipticity of
radiation emitted by accelerated particles can be used to distinguish
electric and toroidal dipole radiation. The work is summarized in
Sec.~\ref{subsec:Conclusion}.

\subsection{Radiation patterns of accelerated point-particle electric, toroidal
and anapole dipoles\label{subsec:RadPat}}
\begin{center}
\begin{figure*}
\includegraphics{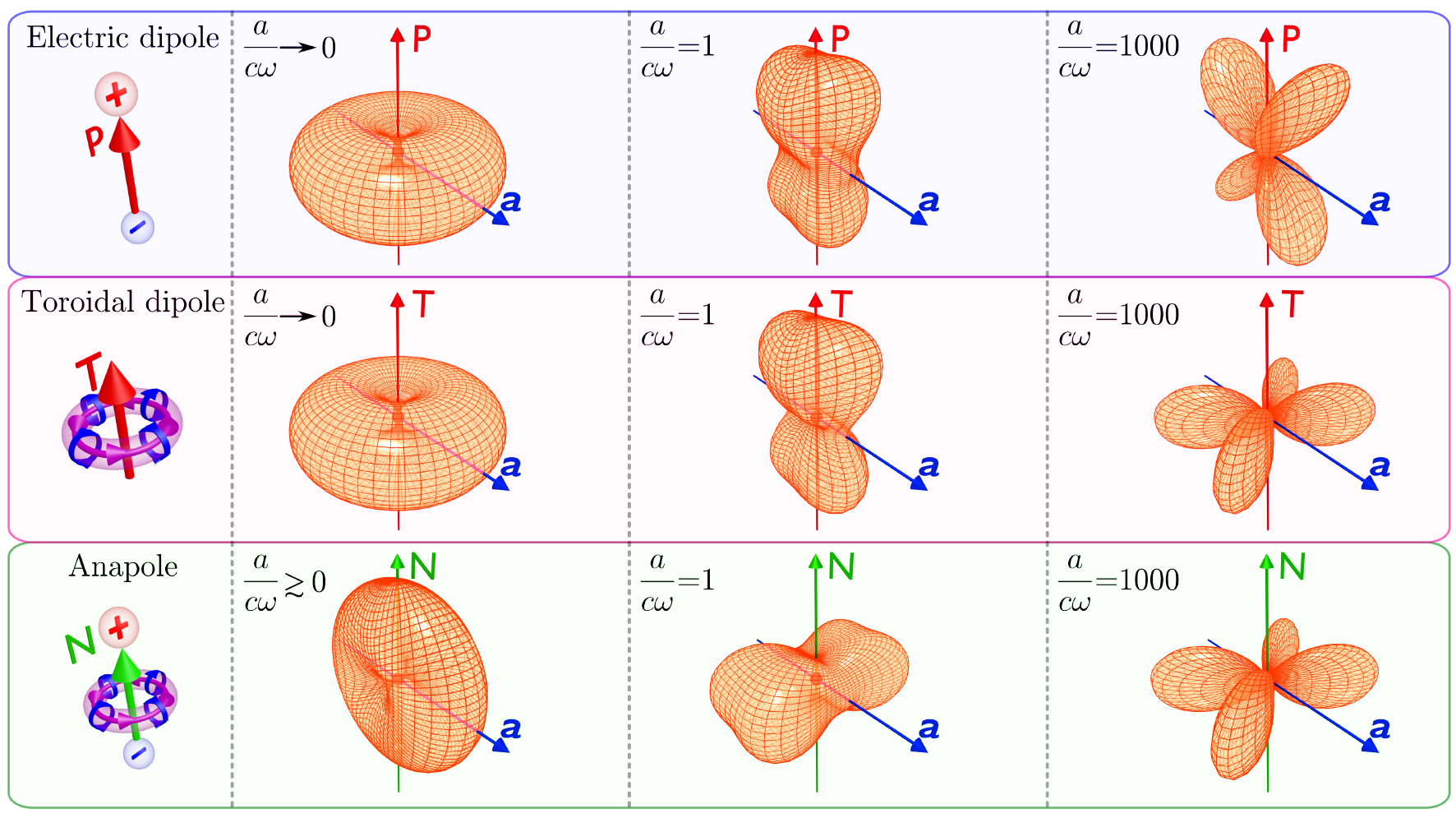}

\caption{\textbf{Radiation patterns of accelerating point-particles with electric
(top row) and toroidal (middle row) and anapole dipoles moments. }The
magnitude of acceleration is $a$. In all cases the corresponding
dipoles are assumed to be harmonically oscillating along a single
axis (e.g. $\vec{p}=\vec{\hat{z}}p_{0}\cos\left(\omega t\right)$
with $p_{0}=const$, and the same for $\vec{T}$ and $\vec{N}$).
The key parameter is $a/\omega c$, where $c$ is the speed of light.
This parameter corresponds to how many cycles of oscillation (i.e.
periods of $2\pi/\omega$) it takes for the accelerating particle
to reach relativistic speed (starting from rest). Three cases are
considered: the 2nd column shows radiation patterns for the case of
low acceleration. For an anapole the acceleration magnitude has to
be above zero for non-zero radiation. The 3rd and 4th columns show
the radiation patterns for the cases of intermediate and high acceleration.
The 4th column also corresponds to the radiation patterns of DC dipoles
(i.e. zero-frequency limit $\omega\to0,\,a>0$). }

\label{fig:AllRadPatterns}
\end{figure*}
\par\end{center}

Given the expressions for the far-field radiation from accelerated
electric, toroidal and anapole dipoles, one can compute their radiation
patterns using the Poynting vector:

\begin{equation}
\vec{S}\left(t,\,\vec{r}\right)=\vec{E}\left(t,\,\vec{r}\right)\times\vec{H}\left(t,\,\vec{r}\right)\asteq\frac{1}{\mu_{0}c}\left|\vec{E}\left(t,\,\vec{r}\right)\right|^{2}\vec{\hat{r}}\label{eq:Main_Poynting}
\end{equation}

The second equality in the equation above applies only in the far-field,
where both electric and magnetic fields are transverse, and where
$\vec{H}\asteq\vec{\hat{r}}\times\vec{E}/\mu_{0}c$. The radiation
patterns of all three dipoles are visualized in Fig.~\ref{fig:AllRadPatterns},
which shows time-averaged power-per-solid angle radiated by accelerated
point-particles with corresponding dipole moments. For the purposes
of visualization, it is assumed that the corresponding dipole moments
oscillate along a single axis with angular frequency $\omega$ (e.g.
$\vec{N}=\vec{\hat{z}}N_{0}\cos\omega t$ for anapole), and that the
acceleration of the particle is perpendicular to the dipole moment
($\vec{N}.\vec{\hat{a}}=0$). In this case, the radiation pattern
is governed by a single parameter $a/\omega c$ which corresponds
to the number of oscillation cycles at frequency $\omega$ it would
take for the particle to reach relativistic speed, starting from rest. 

At low acceleration, the emission of oscillating electric and toroidal
dipoles is indistinguishable, and is in the form of the well-known
doughnut-shaped radiation pattern (see Fig.~\ref{fig:AllRadPatterns}
column 2, rows 1,2). The emission of anapole is more interesting.
As follows from Eq.~(\ref{eq:Main_AnaElecF}), the emission of anapole
vanishes unless: (a)~the anapole is accelerating ($a>0$); (b)~the
anapole moment, or its derivatives, are perpendicular to acceleration
($\vec{\hat{a}}\times\vec{N}\neq0$ or $\vec{\hat{a}}\times\vec{\dot{N}}\neq0$
or $\vec{\hat{a}}\times\vec{\ddot{N}}\neq0$). The low-acceleration
limit of emission from an oscillating anapole point-particle (row
3, column 2 of Fig.~\ref{fig:AllRadPatterns}) has a particularly
simple form, with electric field:
\[
\lim_{a\to0}\vec{E}^{\left(N\right)}\asteq\frac{\mu_{0}}{4\pi rc^{2}}\left(\vec{\hat{r}}\times\vec{a}\times\ddot{\vec{N}}\right)
\]

which corresponds to emission of a \emph{magnetic} point-dipole with
moment $\vec{a}\times\vec{N}/c$. 

As intermediate acceleration ($a/\omega c\sim1$), the radiation pattern
of of all three dipoles becomes significantly more complex (column
3 of Fig.~\ref{fig:AllRadPatterns}). One noteworthy feature is that
in this regime the emission patterns of all three dipoles have no
zeros, i.e. the dipoles emit in all directions with roughly the same
efficiency.

The regime of $a/\omega c\gg1$ (column 4 in Fig.~\ref{fig:AllRadPatterns})
corresponds both to high acceleration of oscillating dipoles, as well
as to emission from static dipoles (where $\omega\to0$). Here the
emission patterns of electric and toroidal dipoles loose any resemblance,
whilst the emission of anapole becomes identical to that of a toroidal
dipole. Indeed, this agrees with the definition of an anapole given
in Eq.~(\ref{eq:Main_AnaMomDef}): static anapole \emph{is} the toroidal
dipole.

\subsection{Distinguishing between electric and toroidal dipoles using ellipticity\label{subsec:Ellipticity} }

The existence of anapoles, and nonradiating configurations more generally,
in electrodynamics is related directly to impossibility to `invert'
the scattering problem, i.e. to uniquely determine the source of electromagnetic
radiation based solely on the electromagnetic (far-)field it emits.
Here we will consider a method of distinguishing between emission
of oscillating electric and toroidal dipole point-particles under
acceleration which relies on the ellipticity of the emitted light.

We consider a scenario in which one analyzes the light scattered (re-emitted)
by atoms via electric and toroidal dipole excitations. It has already
been shown that it is possible to induce both electric \citep{Au78},
and toroidal \citep{Costescu91,Lewis98} dipole excitations in atoms.
However, as discussed in Sec.~\ref{sec:Introduction}, there is currently
no way to experimentally distinguish between them (this is part of
the inverse-source problem). The radiation patterns of accelerated
dipoles in Fig.~\ref{fig:AllRadPatterns} suggest that if the atoms
in question could be subjected to sufficiently high acceleration,
the difference between electric and toroidal dipole excitations would
be detectable. We shall now estimate a realistic value for acceleration
to which the atoms could be subjected. Let the frequency of the light
scattered by the atoms correspond to a free-space wavelength of $\lambda_{0}=1\,\upmu\text{m}$
($\omega=2\pi c/\lambda_{0}$). The atoms under consideration could
be accelerated in a number of ways including sonoluminescence \citep{Weninger97},
and laser based methods \citep{Mourou06,Barker12} which can provide
accelerations as high as $10^{22}\,\text{m/s}^{2}$. The most serious
limitation is that the acceleration should not destroy the atom. The
typical forces within the atom are $F_{0}=10^{-7}\,\text{N}$ (force
on the electron due to the charged nucleus). We shall assume that
the accelerated atom has the mass of a carbon atom ($m_{C}$), and
is accelerated with a force two orders of magnitude smaller than the
internal atomic forces so: $a=F_{0}/\left(100\,m_{C}\right)\approx4\times10^{16}\,\text{m/s}^{2}$.
Therefore $a/\omega c=a\lambda_{0}/2\pi c^{2}\approx10^{-7}$.
\begin{center}
\begin{figure*}
\begin{centering}
\includegraphics{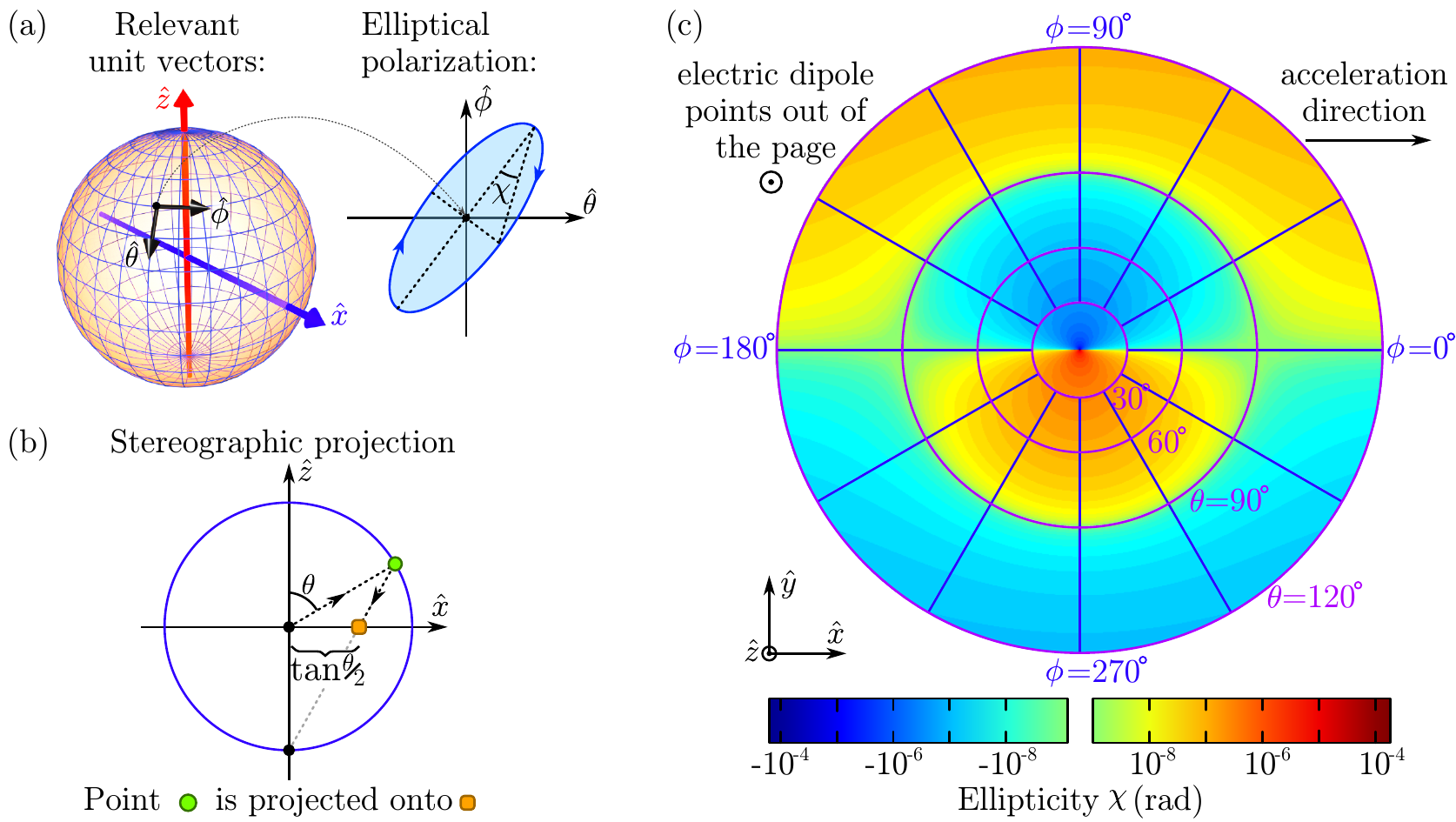}
\par\end{centering}
\caption{\textbf{Ellipticity of the light emitted by electric dipole point-particle.}
\textbf{(a)~Definition of the ellipticity $\chi$.} The radiation
is emitted by a localized particle at the origin. Since polarization
of the emitted radiation is transverse, one can decompose it into
azimuthal ($\vec{\hat{\phi}}$) and polar ($\vec{\hat{\theta}}$)
components. Elliptical polarization corresponds to electric field
of the emitted radiation tracing an ellipse in the $\phi\theta$-plane.
Ellipticity ($\chi$) is defined as the angle, the tangent of which
is the ratio of the semi-minor and semi-major radii of the polarization
ellipse. Positive angle of ellipticity corresponds to electric field
rotating in clock-wise direction (as shown in (a)) around the direction
of radiation propagation. \textbf{(b)~Definition of the stereographic
projection. }The projection allows depicting a scalar field, defined
on a surface of a unit sphere, on a flat plane. Here, for simplicity,
we show a circle instead of the sphere and a line instead of the plane.
A chosen point on the circle (green circle), which corresponds to
angle $\theta$ is projected onto x\textendash axis by plotting a
straight line from the south-pole of the circle to the chosen point.
The point at which the straight line intersects the x-axis corresponds
to the position of the projected point (orange square). Formally,
all points on the surface of a sphere, parametrized by ($\theta,\phi$),
are projected onto points on a plane with Cartesian coordinates $(\tan\frac{\theta}{2}\,\cos\phi,\tan\frac{\theta}{2}\,\sin\phi)$.
\textbf{(c)~Colormap of the ellipticity of the radiation from the
accelerated electric dipole point-particle.} The electric dipole moment
of the particle is $\vec{p}=p_{0}\cos\left(\omega t\right)\vec{\hat{z}}$,
whilst the normalized acceleration is $a/\omega c=10^{-7}$ in the
$\vec{\hat{x}}$-direction. The ellipticity is plotted in a stereographic
projection (see (b)) with polar angle ranging from $\theta=0^{\circ}$
(north-pole) to $\theta=120^{\circ}$. The blue color on the colormap
corresponds to negative ellipticity, and is clipped for $\chi>-10^{-9}\,\text{rad}$.
The red/orange color corresponds to positive ellipticity and is clipped
for $\chi<10^{-9}\,\text{rad}$.}

\label{fig:Ellipticity}
\end{figure*}
\par\end{center}

The the small effects that will arise in the emission patterns of
the atoms subjected to acceleration with $a/\omega c\sim10^{-7}$
are best detected via interference. In particular, here we shall focus
on the ellipticity of the light emitted by accelerated point-particle
dipoles. The definition of the ellipticity $\chi$ is illustrated
in Fig.~\ref{fig:Ellipticity}a. Given the localized electromagnetic
source (at the origin), the polarization of the emitted/scattered
light can be decomposed entirely into azimuthal (unit vector $\vec{\hat{\phi}}$)
and polar (unit vector $\vec{\hat{\theta}}$) components. Treating
the former as `vertical' and the latter as `horizontal' components,
the ellipticity of the emitted light becomes \citep{KligerBook}:
\begin{equation}
\chi=\arcsin\left(\frac{2\,\Im\left(\tilde{E}_{\phi}^{*}\tilde{E}_{\theta}\right)}{\left|\tilde{E}_{\theta}\right|^{2}+\left|\tilde{E}_{\phi}\right|^{2}}\right)/2\label{eq:EllDef}
\end{equation}

Above we have switched to time-harmonic (electric) fields (denoted
by $\tilde{\dots}$) with time-dependence given by $\exp\left(+i\omega t\right)$.
In what is to follow it shall be assumed that the relevant dipoles
are aligned along z-axis and that the acceleration is along x-axis.
The tangent of ellipticity angle $\tan\chi$ is the ratio of the semi-minor
to semi-major radius of the ellipse traced out by the electric field
emitted from the source (see Fig.~\ref{fig:Ellipticity}a). The range
of ellipticity is $-\pi/4\le\chi\le\pi/4$, with two extremes corresponding
to right- ($\chi=\pi/4$) and left-circular ($\chi=-\pi/4$) polarization,
and $\chi=0$ corresponding to linear polarization. 

At zero acceleration the emission of both electric and toroidal dipoles
is linearly polarized (along $\vec{\hat{\theta}}$), the ellipticity
is therefore zero. Acceleration leads to elliptical polarization by
adding a delayed orthogonally polarized electric field to the main
emission pattern. Using Eq.~(\ref{eq:EllDef}) and the time-harmonic
versions of Eqs.~(\ref{eq:Main_EDipElecF},~\ref{eq:Main_TDipElecF}),
the ellipticity of the radiation from accelerated electric ($\chi^{\left(p\right)}$)
and toroidal ($\chi^{\left(T\right)}$) dipoles can be shown to be:
\begin{flalign}
\chi^{\left(p\right)}= & 2\cot(\theta)\sin(\phi)\cdot\frac{a}{\omega c}+\mathcal{O}\left(\left(\frac{a}{\omega c}\right)^{3}\right)\label{eq:EDip_Chi}\\
\chi^{\left(T\right)}= & 3\cot(\theta)\sin(\phi)\cdot\frac{a}{\omega c}+\mathcal{O}\left(\left(\frac{a}{\omega c}\right)^{3}\right)\label{eq:TDip_Chi}
\end{flalign}

Thus, for low accelerations $\chi^{\left(p\right)}/\chi^{\left(T\right)}\approx2/3$.
This is a remarkably simple result considering the complexity of the
underlying radiation patterns. Ellipticity can be visualized by mapping
it onto the unit-sphere and then using stereographic projection to
project a portion of the sphere onto the plane. Stereographic projection
is illustrated in Fig.~\ref{fig:Ellipticity}b. The ellipticity of
the radiation emitted by accelerated electric dipole, with normalized
acceleration $a/\omega c=10^{-7}$, is shown in Fig.~\ref{fig:Ellipticity}c.
Whilst $\chi$ is low at the equatorial plane ($\theta=90^{\circ}$)
where dipole emission is strongest (see Fig.~\ref{fig:AllRadPatterns}),
it can be quite high close to the poles, for example at $\theta=13^{\circ}$
and $\phi=90^{\circ}$ (emitted power 20 times less than at equator)
$\chi=10^{-6}\,\text{rad}$. Equations~(\ref{eq:EDip_Chi}, \ref{eq:TDip_Chi})
suggest that even 10\% sensitivity would be enough to distinguish
between electric and toroidal dipoles, i.e. the required ellipticity
sensitivity is $\Delta\chi\le10^{-7}\,\text{rad}$. Since ellipticity
sensitivity of $\Delta\chi\le10^{-8}\,\text{rad}$ \citep{Carusotto84,Cameron91}
and even $\Delta\chi\le10^{-10}\,\text{rad}$ \citep{Muroo03}, has
already been experimentally demonstrated, we argue that difference
between electric and toroidal dipole excitations in atoms should be
observable with modern technology.

\subsection{Conclusion\label{subsec:Conclusion}}

In conclusion, we have demonstrated that anapoles, elementary non-radiating
configurations, start radiating when accelerated. The consequence
of this loop-hole in the inverse source problem is that one can, in
principle, identify the exact composition of a source of radiation
solely from its emission, as long as the source can be accelerated.
In the process of our analysis we have, for the first time, derived
expressions for the radiation from an accelerated toroidal and anapole
dipoles, and explored polarization properties of the light emitted
by accelerated electric and toroidal dipoles. In particular, we have
demonstrated that ellipticity of the light emitted by small accelerated
particles, such as atoms, can be used to determine whether the emission
is due to electric or toroidal dipole transitions. Our work suggests,
that there are practical ways to circumvent the inverse-source problem.
There are implications for all branches of science where light is
used for characterization, from astronomy to biology and high-energy
physics.
\begin{acknowledgments}
The author acknowledges financial support from the UK Engineering
and Physical Sciences Research Council (Grant No. EP/M009122/1) .
Author also gratefully acknowledges helpful discussions with Prof
N. I. Zheludev (University of Southampton, UK) and Dr V. A. Fedotov
(University of Southampton, UK).
\end{acknowledgments}


\clearpage{}

\appendix

\renewcommand{\thesubsection}{\Alph{subsection}}
\renewcommand{\theequation}{\thesubsection\arabic{equation}}

\section*{Appendix: Light emission by accelerated electric, toroidal and anapole
dipolar sources}

\subsection{Basic conventions/definitions\label{app_subsec:Basics}}

\setcounter{equation}{0}

In order to keep the paper self-contained the basic conventions that
will be used to handle the kinematics of the uniformly accelerated
particles will be given here. 

In all cases we will be working in flat space-time and standard diagonal
metric $g_{\alpha\beta}=diag\left(1,-1,-1,-1\right)$, thus the square
of separation between the two events that differ by time $dt$ and
spatial position $d\vec{r}=\vec{\hat{x}}dx+\vec{\hat{y}}dy+\vec{\hat{z}}dz$
is:
\[
ds^{2}=c^{2}dt^{2}-d\vec{r}.d\vec{r}=c^{2}dt^{2}-dx^{2}-dy^{2}-dz^{2}
\]

The contraction of two four-vectors will be denoted as:
\begin{flalign*}
a\cdot b= & a_{\mu}b^{\mu}=a^{\mu}b_{\mu}=g_{\mu\nu}a^{\mu}b^{\nu}=\\
= & a^{0}b^{0}-a^{1}b^{1}-a^{2}b^{2}-a^{3}b^{3}
\end{flalign*}

Greek indices run through $\left\{ 0,1,2,3\right\} =\left\{ t,x,y,z\right\} $,
and repeated index implies summation unless otherwise stated. 

The three-vectors will be denoted by bold-face font. The contraction
between two three-vectors is denotes as (no summation implied):
\[
\vec{a}.\vec{b}=a^{x}b^{x}+a^{y}b^{y}+a^{z}b^{z}=a^{1}b^{1}+a^{2}b^{2}+a^{3}b^{3}
\]

The partial derivative with respect to coordinate $x^{\alpha}$ is
denoted as $\partial_{\alpha}=\partial/\partial x^{\alpha}$. The
covariant partial derivative along the direction of changing $x^{\alpha}$,
based on Levi-Civita connection, is denoted as $\nabla_{\alpha}$
.

We will be using the standard Kronecker delta tensor \citep{LovelockBook}:
\begin{align*}
\delta_{\beta}^{\alpha}=\frac{\partial x^{\alpha}}{\partial x^{\beta}},\quad & \delta_{\beta}^{\alpha}=g^{\alpha\mu}g_{\mu\beta},\: & \delta_{\beta}^{\alpha}=\begin{cases}
1, & \alpha=\beta\\
0, & otherwise
\end{cases}
\end{align*}

Where $g^{\alpha\beta}$ is the inverse metric. The generalized Kronecker
delta tensor is defined through the determinant \citep{LovelockBook}:
\begin{equation}
\delta_{\beta_{1}\dots\beta_{r}}^{\alpha_{1}\dots\alpha_{r}}=\left|\begin{array}{ccc}
\delta_{\beta_{1}}^{\alpha_{1}} & \dots & \delta_{\beta_{r}}^{\alpha_{1}}\\
\vdots & \ddots & \vdots\\
\delta_{\beta_{1}}^{\alpha_{r}} & \dots & \delta_{\beta_{r}}^{\alpha_{r}}
\end{array}\right|\label{eq:GenKroDef}
\end{equation}

We will also be using Levi-Civita relative tensors \citep{LovelockBook}:
\begin{gather*}
\epsilon^{\alpha\beta\gamma\kappa}=\delta_{0123}^{\alpha\beta\gamma\kappa},\quad\epsilon_{\alpha\beta\gamma\kappa}=\delta_{\alpha\beta\gamma\kappa}^{0123}\\
\delta_{\phi\eta\mu\nu}^{\alpha\beta\gamma\kappa}=\epsilon^{\alpha\beta\gamma\kappa}\epsilon_{\phi\eta\mu\nu}
\end{gather*}

\subsection{Far-field radiation from uniformly accelerating toroidal dipole \label{app_subsec:toroRad}}

\setcounter{equation}{0}

Here we evaluate the electromagnetic tensor for the accelerated toroidal
dipole point-particle. The starting point is Eq.~(\ref{eq:Main_FToro}).
In the first stage we define $Q_{\nu}^{\phi\sigma}=T^{\phi}\hat{\tau}^{\sigma}\hat{\tau}_{\nu}$
and differentiate the $1/\left(u\cdot X\right)$ terms:

\begin{widetext}
\begin{flalign}
F_{\mu\eta}^{\left(T\right)}\doteq & \frac{\mu_{0}c^{2}}{4\pi}\delta_{\mu\eta}^{\theta\xi}\,g_{\xi\kappa}\delta_{\phi\sigma\rho}^{\kappa\nu\alpha}g^{\rho\beta}\left[\frac{1}{u\cdot X}\,\frac{d}{d\tau}\left[\left(\frac{X_{\theta}}{u\cdot X}\right)\frac{d}{d\tau}\left[\left(\frac{X_{\alpha}}{u\cdot X}\right)\frac{d}{d\tau}\left[\left(\frac{X_{\beta}}{u\cdot X}\right)T^{\phi}\hat{\tau}^{\sigma}\hat{\tau}_{\nu}\right]\right]\right]\right]_{\tau=\tau_{ret}}\nonumber \\
\doteq & \frac{\mu_{0}c^{2}}{4\pi}\delta_{\mu\eta}^{\theta\xi}\,g_{\xi\kappa}\delta_{\phi\sigma\rho}^{\kappa\nu\alpha}g^{\rho\beta}\left[\frac{X_{\theta}X_{\alpha}X_{\beta}}{\left(u\cdot X\right)^{4}}\left(\frac{d^{3}Q_{\nu}^{\phi\sigma}}{d\tau^{3}}+\left(-6\frac{a\cdot X}{u\cdot X}\right)\frac{d^{2}Q_{\nu}^{\phi\sigma}}{d\tau^{2}}+\right.\right.\nonumber \\
 & \left.\left.+\left(15\left(\frac{a\cdot X}{u\cdot X}\right)^{2}-4\left(\frac{a}{c}\right)^{2}\right)\frac{dQ_{\nu}^{\phi\sigma}}{d\tau}+\left(9\left(\frac{a}{c}\right)^{2}\left(\frac{a\cdot X}{u\cdot X}\right)-15\left(\frac{a\cdot X}{u\cdot X}\right)^{3}\right)Q_{\nu}^{\phi\sigma}\right)\right]_{\tau=\tau_{ret}}\label{eq:TDipFQ}
\end{flalign}

\end{widetext}

Next, one evaluates the derivatives:

\begin{widetext}

\begin{flalign}
\frac{dQ_{\nu}^{\phi\sigma}}{d\tau}= & \left(\dot{T}^{\phi}\right)\hat{\tau}^{\sigma}\hat{\tau}_{\nu}+\left(\frac{a}{c}T^{\phi}\right)\hat{a}^{\sigma}\hat{\tau}_{\nu}+\left(\frac{a}{c}T^{\phi}\right)\hat{\tau}^{\sigma}\hat{a}_{\nu}\nonumber \\
\frac{d^{2}Q_{\nu}^{\phi\sigma}}{d\tau^{2}}= & \left(\ddot{T}^{\phi}+2\left(\frac{a}{c}\right)^{2}T^{\phi}\right)\hat{\tau}^{\sigma}\hat{\tau}_{\nu}+\left(2\left(\frac{a}{c}\right)\dot{T}^{\phi}\right)\hat{a}^{\sigma}\hat{\tau}_{\nu}+\left(2\left(\frac{a}{c}\right)\dot{T}^{\phi}\right)\hat{\tau}^{\sigma}\hat{a}_{\nu}+\left(2\left(\frac{a}{c}\right)^{2}T^{\phi}\right)\hat{a}^{\sigma}\hat{a}_{\nu}+\nonumber \\
 & \quad+\left(-\left(\frac{a}{c}\right)^{2}\left(\hat{a}\cdot T\right)\right)\hat{\tau}^{\phi}\hat{a}^{\sigma}\hat{\tau}_{\nu}\nonumber \\
\frac{d^{3}Q_{\nu}^{\phi\sigma}}{d\tau^{3}}= & \left(\dddot{T}^{\phi}+6\left(\frac{a}{c}\right)^{2}\dot{T}^{\phi}\right)\hat{\tau}^{\sigma}\hat{\tau}_{\nu}+\left(3\left(\frac{a}{c}\right)\ddot{T}^{\phi}+4\left(\frac{a}{c}\right)^{3}T^{\phi}\right)\hat{a}^{\sigma}\hat{\tau}_{\nu}+\left(3\left(\frac{a}{c}\right)\ddot{T}^{\phi}+4\left(\frac{a}{c}\right)^{3}T^{\phi}\right)\hat{\tau}^{\sigma}\hat{a}_{\nu}\nonumber \\
 & \quad+\left(6\left(\frac{a}{c}\right)^{2}\dot{T}^{\phi}\right)\hat{a}^{\sigma}\hat{a}_{\nu}+\left(-3\left(\frac{a}{c}\right)^{2}\left(\hat{a}\cdot\dot{T}\right)\right)\hat{\tau}^{\phi}\hat{a}^{\sigma}\hat{\tau}_{\nu}+\left(-3\left(\frac{a}{c}\right)^{3}\left(\hat{a}\cdot T\right)\right)\hat{\tau}^{\phi}\hat{a}^{\sigma}\hat{a}_{\nu}\label{eq:TDipQDeriv}
\end{flalign}

\end{widetext}

Above we have used the rule $dT^{\phi}/d\tau=\dot{T}^{\phi}-\frac{a}{c}\left(\hat{a}\cdot T\right)\hat{\tau}^{\phi}$,
where $\dot{T}^{\phi}=\frac{dT^{1}}{d\tau}\hat{x}^{\phi}+\frac{dT^{2}}{d\tau}\hat{y}^{\phi}+\frac{dT^{a}}{d\tau}\hat{a}^{\phi}$,
and the second term in the derivative arises due to $d\hat{a}^{\phi}/d\tau=\left(a/c\right)\hat{\tau}^{\phi}$.
However, in the presence of $\delta_{\phi\sigma\rho}^{\mu\nu\alpha}\hat{\tau}^{\sigma}$,
this rule can be simplified to $\delta_{\phi\sigma\rho}^{\mu\nu\alpha}\left(dT^{\phi}/d\tau\right)\hat{\tau}^{\sigma}=\delta_{\phi\sigma\rho}^{\mu\nu\alpha}\dot{T}^{\phi}\hat{\tau}^{\sigma}$
because of the anti-symmetry of $\delta_{\phi\sigma\rho}^{\mu\nu\alpha}$.
Combining Eq.~(\ref{eq:TDipFQ}) and Eq.~(\ref{eq:TDipQDeriv}):

\begin{widetext}

\begin{flalign*}
F_{\mu\eta}^{\left(T\right)}\doteq & \frac{\mu_{0}c^{2}}{4\pi}\delta_{\mu\eta}^{\theta\xi}\,g_{\xi\kappa}\delta_{\phi\sigma\rho}^{\kappa\nu\alpha}g^{\rho\beta}\left[\frac{X_{\theta}X_{\alpha}X_{\beta}}{\left(u\cdot X\right)^{4}}\Biggl\{\right.\\
 & \,\,\,\left(\dddot{T}^{\phi}-6\left(\frac{a\cdot X}{u\cdot X}\right)\ddot{T}^{\phi}+\left(2\left(\frac{a}{c}\right)^{2}+15\left(\frac{a\cdot X}{u\cdot X}\right)^{2}\right)\dot{T}^{\phi}-\left(3\left(\frac{a}{c}\right)^{2}\left(\frac{a\cdot X}{u\cdot X}\right)+15\left(\frac{a\cdot X}{u\cdot X}\right)^{3}\right)T^{\phi}\right)\hat{\tau}^{\sigma}\hat{\tau}_{\nu}+\\
 & +3\left(\frac{a}{c}\right)\left(\ddot{T}^{\phi}-4\left(\frac{a\cdot X}{u\cdot X}\right)\dot{T}^{\phi}+5\left(\frac{a\cdot X}{u\cdot X}\right)^{2}T^{\phi}\right)\left(\hat{a}^{\sigma}\hat{\tau}_{\nu}+\hat{\tau}^{\sigma}\hat{a}_{\nu}\right)+6\left(\frac{a}{c}\right)^{2}\left(\dot{T}^{\phi}-2\left(\frac{a\cdot X}{u\cdot X}\right)T^{\phi}\right)\hat{a}^{\sigma}\hat{a}_{\nu}-\\
 & \left.-3\left(\frac{a}{c}\right)^{2}\left(\left(\hat{a}\cdot\dot{T}\right)-2\left(\frac{a\cdot X}{u\cdot X}\right)\left(\hat{a}\cdot T\right)\right)\hat{\tau}^{\phi}\hat{a}^{\sigma}\hat{\tau}_{\nu}-3\left(\frac{a}{c}\right)^{3}\left(\hat{a}\cdot T\right)\hat{\tau}^{\phi}\hat{a}^{\sigma}\hat{a}_{\nu}\Biggr\}\right]
\end{flalign*}

\end{widetext}

As with electric dipole, we approximate the above equation for the
case when speed of the toroidal dipole, in the lab-frame, is insignificantly
small. The necessary transformations are in Eq.~(\ref{eq:Main_LabFrameApprox})
and:
\begin{align*}
T^{\mu}\to & \left(0,\vec{T}\right)^{\mu} & \hat{\tau}_{\mu}\to & \delta_{\mu}^{0}\\
\hat{a}\cdot T\to & -\hat{\vec{a}}.\vec{T} & \hat{a}_{\mu}\to & -\delta_{\mu}^{3}
\end{align*}

The electromagnetic tensor of the point-particle with toroidal dipole,
that is momentarily at rest in the lab frame then becomes:

\begin{widetext}

\begin{flalign*}
F_{\mu\eta}^{\left(T\right)}\asteq & \frac{\mu_{0}}{4\pi rc^{2}}\,\delta_{\mu\eta}^{\theta\xi}\left(1,\,\vec{\hat{r}}\right)_{\theta}g_{\xi\kappa}\,\delta_{\phi\sigma\beta}^{\kappa\nu\alpha}\left(1,\,\vec{\hat{r}}\right)_{\alpha}\left(1,\,\vec{\hat{r}}\right)^{\beta}\Biggl\{\\
 & \,\,\,\left(0,\,\vec{\mathcal{K}}\right)^{\phi}\delta_{0}^{\sigma}\delta_{\nu}^{0}+3\left(\frac{a}{c}\right)\left(0,\,\vec{\mathcal{L}}\right)^{\phi}\left(\delta_{3}^{\sigma}\delta_{\nu}^{0}-\delta_{0}^{\sigma}\delta_{\nu}^{3}\right)-6\left(\frac{a}{c}\right)^{2}\left(0,\,\vec{\mathcal{M}}\right)^{\phi}\delta_{3}^{\sigma}\delta_{\nu}^{3}+3\left(\frac{a}{c}\right)^{2}\left(\vec{\hat{a}}.\vec{\mathcal{M}}\right)\delta_{0}^{\phi}\delta_{3}^{\sigma}\delta_{\nu}^{0}-\\
 & -3\left(\frac{a}{c}\right)^{3}\left(\vec{\hat{a}}.\vec{T}\right)\delta_{0}^{\phi}\delta_{3}^{\sigma}\delta_{\nu}^{3}\Biggr\}\\
\vec{\mathcal{K}}= & \dddot{\vec{T}}+6\left(\frac{a}{c}\right)\left(\vec{\hat{a}}.\vec{\hat{r}}\right)\ddot{\vec{T}}+\left(\frac{a}{c}\right)^{2}\left(2+15\left(\vec{\hat{a}}.\vec{\hat{r}}\right)^{2}\right)\dot{\vec{T}}+3\left(\frac{a}{c}\right)^{3}\left(\vec{\hat{a}}.\vec{\hat{r}}\right)\left(1+5\left(\vec{\hat{a}}.\vec{\hat{r}}\right)^{2}\right)\vec{T}\\
\vec{\mathcal{L}}= & \ddot{\vec{T}}+4\left(\frac{a}{c}\right)\left(\vec{\hat{a}}.\vec{\hat{r}}\right)\dot{\vec{T}}+5\left(\frac{a}{c}\right)^{2}\left(\vec{\hat{a}}.\vec{\hat{r}}\right)^{2}\vec{T}\\
\vec{\mathcal{M}}= & \dot{\vec{T}}+2\left(\frac{a}{c}\right)\left(\vec{\hat{a}}.\vec{\hat{r}}\right)\vec{T}
\end{flalign*}

\end{widetext}

Above the, vectors $\vec{\mathcal{K}}$, $\vec{\mathcal{L}}$ , and
$\vec{\mathcal{M}}$ were introduced to simplify and shorten expression.
Next, we find the electric ($\left(\vec{E}\right)^{i}=-g^{is}cF_{0s}$)
and magnetic fields ($\left(\mu_{0}\vec{H}\right)^{i}=-\epsilon^{iab}F_{ab}/2$): 

\begin{widetext}

\begin{flalign*}
\vec{E}^{\left(T\right)}\asteq & \frac{\mu_{0}}{4\pi rc}\left[-\vec{\hat{r}}\times\vec{\hat{r}}\times\vec{\mathcal{K}}+3\left(\frac{a}{c}\right)\vec{\hat{r}}\times\vec{\hat{a}}\times\vec{\mathcal{L}}+3\left(\frac{a}{c}\right)\left(\vec{\hat{a}}.\vec{\hat{r}}\right)\vec{\hat{r}}\times\vec{\hat{r}}\times\vec{\mathcal{L}}-6\left(\frac{a}{c}\right)^{2}\left(\vec{\hat{a}}.\vec{\hat{r}}\right)\vec{\hat{r}}\times\vec{\hat{a}}\times\vec{\mathcal{M}}+\right.\\
 & \left.\quad\quad\quad+3\left(\frac{a}{c}\right)^{2}\left(\vec{\hat{a}}.\vec{\mathcal{M}}\right)\vec{\hat{r}}\times\vec{\hat{r}}\times\vec{\hat{a}}-3\left(\frac{a}{c}\right)^{3}\left(\vec{\hat{a}}.\vec{\hat{r}}\right)\left(\vec{\hat{a}}.\vec{T}\right)\vec{\hat{r}}\times\vec{\hat{r}}\times\vec{\hat{a}}\right]_{\vec{T}=\vec{T}\left(t-r/c\right)}\\
\vec{H}^{\left(T\right)}\asteq & \frac{1}{4\pi rc^{2}}\left[\vec{\hat{r}}\times\vec{\mathcal{K}}+3\left(\frac{a}{c}\right)\vec{\hat{r}}\times\vec{\hat{r}}\times\vec{\hat{a}}\times\vec{\mathcal{L}}-3\left(\frac{a}{c}\right)\left(\vec{\hat{a}}.\vec{\hat{r}}\right)\vec{\hat{r}}\times\vec{\mathcal{L}}-6\left(\frac{a}{c}\right)^{2}\left(\vec{\hat{a}}.\vec{\hat{r}}\right)\vec{\hat{r}}\times\vec{\hat{r}}\times\vec{\hat{a}}\times\vec{\mathcal{M}}-\right.\\
 & \left.\quad\quad\quad-3\left(\frac{a}{c}\right)^{2}\left(\vec{\hat{a}}.\vec{\mathcal{M}}\right)\vec{\hat{r}}\times\vec{\hat{a}}+3\left(\frac{a}{c}\right)^{3}\left(\vec{\hat{a}}.\vec{\hat{r}}\right)\left(\vec{\hat{a}}.\vec{T}\right)\vec{\hat{r}}\times\vec{\hat{a}}\right]_{\vec{T}=\vec{T}\left(t-r/c\right)}
\end{flalign*}

\end{widetext}

The expressions above can be recombined and shortened. To this end
we introduce the following vectors:

\begin{flalign*}
\vec{\mathcal{F}}= & \dddot{\vec{T}}+3\left(\frac{a}{c}\right)\left(\vec{\hat{a}}.\vec{\hat{r}}\right)\ddot{\vec{T}}+\left(\frac{a}{c}\right)^{2}\left(2+3\left(\vec{\hat{a}}.\vec{\hat{r}}\right)^{2}\right)\dot{\vec{T}}+\\
 & \quad+3\left(\frac{a}{c}\right)^{3}\left(\vec{\hat{a}}.\vec{\hat{r}}\right)\vec{T}\\
\vec{\mathcal{Q}}= & \ddot{\vec{T}}+2\left(\frac{a}{c}\right)\left(\vec{\hat{a}}.\vec{\hat{r}}\right)\dot{\vec{T}}+\left(\frac{a}{c}\right)^{2}\left(\vec{\hat{a}}.\vec{\hat{r}}\right)^{2}\vec{T}\\
\vec{\mathcal{G}}= & \dot{\vec{T}}+\left(\frac{a}{c}\right)\left(\vec{\hat{a}}.\vec{\hat{r}}\right)\vec{T}
\end{flalign*}

The electric ($\vec{E}^{\left(T\right)}$) and magnetic fields ($\vec{H}^{\left(T\right)}$)
of the point-like particle, with toroidal dipole moment $\vec{T}$,
that is momentarily at rest (at the origin) in the lab frame, but
is undergoing constant acceleration $\vec{a}$, then become:
\begin{flalign}
\vec{E}^{\left(T\right)}\asteq & \frac{\mu_{0}}{4\pi rc}\left[-\vec{\hat{r}}\times\vec{\hat{r}}\times\vec{\mathcal{F}}+3\left(\frac{a}{c}\right)\vec{\hat{r}}\times\vec{\hat{a}}\times\vec{\mathcal{Q}}+\right.\nonumber \\
 & \left.\quad+3\left(\frac{a}{c}\right)^{2}\left(\vec{\hat{a}}.\vec{\mathcal{G}}\right)\vec{\hat{r}}\times\vec{\hat{r}}\times\vec{\hat{a}}\right]_{\vec{T}=\vec{T}\left(t-r/c\right)}\label{eq:TDipElecF}\\
\vec{H}^{\left(T\right)}\asteq & \frac{1}{4\pi rc^{2}}\left[\vec{\hat{r}}\times\vec{\mathcal{F}}+3\left(\frac{a}{c}\right)\vec{\hat{r}}\times\vec{\hat{r}}\times\vec{\hat{a}}\times\vec{\mathcal{Q}}-\right.\nonumber \\
 & \left.\quad-3\left(\frac{a}{c}\right)^{2}\left(\vec{\hat{a}}.\vec{\mathcal{G}}\right)\vec{\hat{r}}\times\vec{\hat{a}}\right]_{\vec{T}=\vec{T}\left(t-r/c\right)}\label{eq:TDipMagF}
\end{flalign}

As in the case of electric dipole, the fields of the toroidal dipole
are transverse, and are related through $\vec{H}^{\left(T\right)}=\frac{1}{\mu_{0}c}\vec{\hat{r}}\times\vec{E}^{\left(T\right)}$
indicating radial Poynting vector.

\subsection{Difference between a moving particle with (lab-frame) toroidization
and particle with toroidal dipole\label{app_subsec:MovingToro}}

In the main text it is stated that properties of non-inertial particles
with net toroidal dipole moment have so far not been analyzed. Here
we clarify this assertion. Work by J.~A.~Heras \citep{Hera98},
has claimed to examine properties of non-inertial particles with toroidal
dipole moment, but in fact considered properties of non-inertial particles
which had net toroidization in the lab-frame. In this section we will
briefly explain the important difference between a particle with toroidal
dipole in its rest-frame, and a particle that appears to have net
toroidal dipole in the lab-frame, i.e. some inertial frame around
which we chose to base our calculations. 

Clearly if a particle with toroidal dipole moment, described by current
density in Eq.~(\ref{eq:Main_TDipJ}), is instantaneously at rest
in the lab-frame, its current density is going to be that of a particle
with toroidal dipole. It is tempting to assume that radiation produced
by a particle with toroidal dipole, when it is instantaneously at
rest in the lab frame, can be obtained by analyzing a particle with
only the toroidization in the lab-frame, i.e. by using equations $\rho=0$
(charge density) and $\vec{J}=\nabla\times\nabla\times c\vec{T}\delta^{\left(3\right)}$,
as it was done in Ref.~\citep{Hera98}. This, however, is incorrect.
Indeed, this is stated in Ref.~\citep{Hera98}.

As is shown in Eq.~(\ref{eq:TDipFQ}), finding the far-field produced
by a particle with toroidal dipole moment requires taking three derivatives
with respect to proper time. Therefore, the instantaneous value of
current density is insufficient to obtain radiation, instead one has
to work with the full expression for the current density (Eq.~(\ref{eq:Main_TDipJ})),
and take the limit of particle being at rest in the lab frame \emph{after}
the electromagnetic fields have been found. Failing to do so results
in some terms of the radiation field being lost without any justification.
The origin of error is failure to take the derivatives of the basis
vectors $\hat{\tau}^{\mu}=u^{\mu}/c$ ($u^{\mu}$ is the four-velocity
of the particle) and $\hat{a}^{\mu}=a^{\mu}/\sqrt{-a\cdot a}$ ($a^{\mu}$
is the four-acceleration of the particle).

\subsection{Inertial motion of an anapole\label{app_subsec:IntertialAna}}

The main focus of this paper is on anapole particles in non-inertial
motion. It is natural to ask what happens in case of inertial motion.
Here we briefly show that anapole in inertial motion does not radiate. 

Given an anapole in inertial motion, one can start with its radiation
in anapole particle's rest frame $\tilde{S}$. It follows from Eq.~(\ref{eq:Main_AnaElecF},\ref{eq:Main_AnaMagF}),
after substitution $a\to0$ for inertial motion, that a non-accelerating
anapole will emit no radiation in its rest frame (indeed this is the
basic property of anapoles \citep{ToroRev16}). Therefore the corresponding
electromagnetic tensor $\tilde{F}_{\mu\eta}=0$. However since Maxwell
Equations are invariant under (static) Lorentz transformation one
can find the radiation in any other inertial frame, including lab-frame
$S$, following:
\[
F_{\mu\eta}=\frac{\partial\tilde{x}^{\alpha}}{\partial x_{\mu}}\cdot\frac{\partial\tilde{x}^{\beta}}{\partial x_{\eta}}\cdot\tilde{F}_{\alpha\beta}=0
\]

The electromagnetic tensor $F_{\mu\eta}$ contains all components
of electric and magnetic field, so if $F_{\mu\eta}=0$ then there
is no radiation in the lab-frame. Therefore, anapoles in inertial
motion do not radiate. 
\end{document}